\documentclass[lettersize,journal]{IEEEtran}
\usepackage{array}
\usepackage[caption=false,font=normalsize,labelfont=sf,textfont=sf]{subfig}
\usepackage{textcomp}
\usepackage{stfloats}
\usepackage{url}
\usepackage{verbatim}
\usepackage{graphicx}
\usepackage{cite}
\usepackage{mathptmx}
\usepackage{amsmath,amssymb,amsfonts}
\usepackage{multicol}        % used for the two-column index
\usepackage{multirow}
\usepackage[noend]{algpseudocode}
\usepackage{verbatim}
\usepackage{bm,epstopdf}

\usepackage{booktabs}
\usepackage{tabularx}
\usepackage{footnote}
\usepackage{lineno}   %使用行号
\usepackage{tikz}						
\usetikzlibrary{positioning} 
\usepackage{bm}
\usepackage{float}
\usepackage{stfloats}
\usetikzlibrary{perspective} % LaTeX and plain TeX
\usetikzlibrary{quotes} % LaTeX and plain TeX
\usetikzlibrary{calc}
\usetikzlibrary{arrows.meta}
\usepackage{algorithm}
\usepackage{array}
\usepackage[caption=false,font=normalsize,labelfont=sf,textfont=sf]{subfig}
\usepackage{textcomp}
\usepackage{url}
\usepackage{cite}

\newtheorem{definition}{Definition}

\newcommand{\T}{\mathrm{T}}
\newcommand{\F}{\mathrm{F}}

\begin{document}
	
	\title{Federated Learning Using Coupled Tensor Train Decomposition}
	
	\author{Xiangtao Zhang,
		Eleftherios Kofidis,~\IEEEmembership{Member,~IEEE}, Ce Zhu,~\IEEEmembership{Fellow,~IEEE}, Le Zhang,~\IEEEmembership{Member,~IEEE} and Yipeng Liu,~\IEEEmembership{Senior Member,~IEEE}
		% <-this % stops a space
		\thanks{X. Zhang, C. Zhu, L. Zhang, and Y. Liu are with the School of Information and Communication Engineering, University of Electronic Science and Technology, Chengdu, Sichuan, China. E-mail: yipengliu@uestc.edu.cn.\\
			\indent E. Kofidis is with the Department of Statistics and Insurance Science, University of Piraeus, Greece, and the Computer Technology Institute \& Press ``Diophantus" (CTI), Greece. Email: kofidis@unipi.gr.}
	}

	
	\maketitle
	
	\begin{abstract}
		%Federated tensor decomposition (FTD) has gained substantial traction as a technique to distill meaningful insights from expansive datasets while upholding data confidentiality. Within this context, coupled tensor decomposition (CTD) emerges as a favored avenue for extracting joint features from multimodal data, effectively presenting a unified lens for comprehending and modeling the intricacies intrinsic to FTD predicaments. 
		Coupled tensor decomposition (CTD) can extract joint features from multimodal data in various applications. It can be employed for federated learning networks with data confidentiality. Federated CTD achieves data privacy protection by sharing common features and keeping individual features. However, traditional CTD schemes based on canonical polyadic decomposition (CPD) may suffer from low computational efficiency and heavy communication costs. Inspired by the efficient tensor train decomposition, we propose a coupled tensor train (CTT) decomposition for federated learning. The distributed coupled multi-way data are decomposed into a series of tensor trains with shared factors. In this way, we can extract common features of coupled modes while maintaining the different features of uncoupled modes. Thus the privacy preservation of information across different network nodes can be ensured. The proposed CTT approach is instantiated for two fundamental network structures, namely master-slave and decentralized networks. Experimental results on synthetic and real datasets demonstrate the superiority of the proposed schemes over existing methods in terms of both computational efficiency and communication rounds. In a classification task, experimental results show that the CTT-based federated learning achieves almost the same accuracy performance as that of the centralized counterpart.
  
	\end{abstract}
	
	\begin{IEEEkeywords}
		coupled tensor decomposition, federated learning, distributed factorization, tensor train. 
	\end{IEEEkeywords}
	
	\section{Introduction}
	\label{sec:intro}
	
	\IEEEPARstart{F}{e}derated learning (FL), as a privacy-preserving distributed machine learning technique~\cite{gafni2022,zhou2023federated}, was initially introduced in a master-slave form~\cite{mcmahan2017communication,sattler2020robust,wang2019adaptive}, where the clients perform local model training and the central server aggregates the training model parameters without the need to bring the data into the central storage. However, as the number of clients increases, the communication cost for the server becomes relatively high~\cite{lian2017can}. To address this issue, decentralized FL~\cite{lu2020decentralized,liu2022decentralized} has been proposed instead. By eliminating the need for a central server~\cite{lu2020decentralized}, neighboring clients can exchange local models to achieve model consensus while preserving privacy and achieving relative robustness to failures~\cite{vanhaesebrouck2017decentralized}. Both these two kinds of FL are considered in the present work, following a coupled tensor decomposition-based approach. 
	
\begin{figure}
		\centering
		\begin{tikzpicture}[scale=0.7]
			\definecolor{Color1}{RGB}{114,188,213}
			\definecolor{Color2}{RGB}{245,178,135}
			\fill[Color1] (1,0,0) -- (1,2,0) -- (3,2,0) -- (3,0,0) -- cycle;
			\fill[Color1] (1,0,2) -- (1,2,2) -- (3,2,2) -- (3,0,2) -- cycle;
			\fill[Color1] (1,0,0) -- (1,0,2)-- (1,2,2)-- (1,2,0)-- cycle;
			\fill[Color1] (1,2,0) -- (1,2,2)-- (3,2,2)-- (3,2,0)-- cycle;
			\fill[Color1] (3,2,0) -- (3,2,2)-- (3,0,2)-- (3,0,0)-- cycle;
			\fill[Color1] (3,0,0) -- (3,0,2) -- (1,0,2) -- (1,0,0)-- cycle;
			\draw[] (1,0,2) -- node [left]{$I$}(1,2,2) -- (3,2,2) -- (3,0,2) --node [below]{$J$} cycle;
			\node (circle) at (2,1,2){$\mathcal{X}$};
			\draw[] (1,2,0) --  node [left,above]{$K$}(1,2,2);
			\draw[] (3,2,0) -- (3,2,2);
			\draw[] (3,0,0) -- (3,0,2);
			\draw[]  (1,2,0) -- (3,2,0);

			\fill[Color2] (3,0,-2) -- (3,2,-2) -- (5.5,2,-2) -- (5.5,0,-2) -- cycle;
			\fill[Color2] (3,0,0) -- (3,2,0) -- (5.5,2,0) -- (5.5,0,0) -- cycle;
			\fill[Color2] (3,0,-2) -- (3,0,0)-- (3,2,0)-- (3,2,-2)-- cycle;
			\fill[Color2] (3,2,-2) -- (3,2,0)-- (5.5,2,0) -- (5.5,2,-2)-- cycle;
			\fill[Color2] (5.5,2,-2) -- (5.5,2,0)-- (5.5,0,0)-- (5.5,0,-2)-- cycle;
			\fill[Color2] (5.5,0,-2) -- (5.5,0,0)--(3,0,0)--(3,0,-2) -- cycle;
			\draw[] (3,0,0) -- (3,2,0) -- (5.5,2,0) --node [right]{$I$} (5.5,0,0) --node [below]{$M$} cycle;
			\node (circle) at (4.25,1,0){$\mathcal{Y}$};
			\draw[] (3,2,-2) --  node [left,above]{$L$}(3,2,0);
			\draw[] (5.5,0,-2) -- (5.5,0,0);
			\draw[] (5.5,2,-2) -- (5.5,2,0);
			\draw[] (3,2,-2) -- (5.5,2,-2);
			\draw[] (5.5,0,-2) -- (5.5,2,-2);
			
		\end{tikzpicture}
		\begin{tikzpicture}[scale=0.7]
			\definecolor{Color1}{RGB}{114,188,213}
			\definecolor{Color2}{RGB}{245,178,135}
			\fill[Color1] (1,0,0) -- (1,2,0) -- (3,2,0) -- (3,0,0) -- cycle;
			\fill[Color1] (1,0,2) -- (1,2,2) -- (3,2,2) -- (3,0,2) -- cycle;
			\fill[Color1] (1,0,0) -- (1,0,2)-- (1,2,2)-- (1,2,0)-- cycle;
			\fill[Color1] (1,2,0) -- (1,2,2)-- (3,2,2)-- (3,2,0)-- cycle;
			\fill[Color1] (3,2,0) -- (3,2,2)-- (3,0,2)-- (3,0,0)-- cycle;
			\fill[Color1] (3,0,0) -- (3,0,2) -- (1,0,2) -- (1,0,0)-- cycle;
			\draw[] (1,0,2) -- node [left]{$I$}(1,2,2) -- (3,2,2) -- (3,0,2) -- node [below]{$J$}cycle;
			\node (circle) at (2,1,2){$\mathcal{X}$};
			\draw[] (1,2,0) --  node [left,above]{$K$} (1,2,2);
			\draw[] (3,2,0) -- (3,2,2);
			\draw[] (3,0,0) -- (3,0,2);
			\draw[]  (1,2,0) -- (3,2,0);

			\fill[Color2] (3,0,0) -- (3,2,0) -- (5.5,2,0) -- (5.5,0,0) -- cycle;
			\fill[Color2] (3,0,2) -- (3,2,2) -- (5.5,2,2) -- (5.5,0,2) -- cycle;
			\fill[Color2] (3,0,0) -- (3,0,2)-- (3,2,2)-- (3,2,0)-- cycle;
			\fill[Color2] (3,2,0) -- (3,2,2)-- (5.5,2,2) -- (5.5,2,0)-- cycle;
			\fill[Color2] (5.5,2,0) -- (5.5,2,2)-- (5.5,0,2)-- (5.5,0,0)-- cycle;
			\fill[Color2] (5.5,0,0) -- (5.5,0,2)--(3,0,2)--(3,0,0) -- cycle;
			\draw[] (3,0,2) -- (3,2,2) -- (5.5,2,2) -- (5.5,0,2) -- node [below]{$M$} cycle;
			\node (circle) at (4.25,1,2){$\mathcal{Y}$};
			\draw[] (3,2,0) -- (3,2,2);
			\draw[] (5.5,0,0) -- (5.5,0,2);
			\draw[] (5.5,2,0) -- (5.5,2,2);
			\draw[] (3,2,0) -- (5.5,2,0);
			\draw[] (5.5,0,0) -- (5.5,2,0);
		\end{tikzpicture}
		\caption{3rd-order tensor coupling across the first mode (left) and across the first and second modes (right).}
		\label{fig:coupled}
	\end{figure}
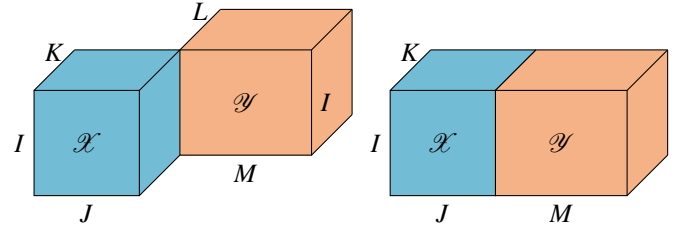

	Higher-order tensors, as a natural extension of matrices in multi-dimensional spaces, have proved to be a natural and powerful tool for modeling~\cite{koniusz2021tensor,wang2022sparse} and extracting features from data with multiple attributes~\cite{sdfhpf17,liu2022,gao2021federated}. Tensor decomposition (TD) models and methods have shown significant potential in numerous data processing~\cite{ho2014limestone,cichocki2015tensor} and analysis problems and have been increasingly adopted in a wide range of application areas~\cite{ong2021protecting,wang2023guaranteed,liu2022}.
	Moreover, multi-way data originating from diverse sources often share underlying information alongside common dimensions/modes~\cite{chatz2022} and can thus be represented as \emph{coupled} tensors~\cite{chatzichristos2022coupled,guo2022multispectral}, as illustrated in Fig.~\ref{fig:coupled}. 
	For instance, electronic health records (EHRs) derived from multiple hospitals can be effectively modeled as coupled tensors, wherein common features such as medical examination results and prescription drugs are shared~\cite{ho2014marble}, while each source retains its specific features related to patient information. By jointly analyzing data from multiple sources, coupled tensor decomposition (CTD) enables the extraction of both shared and individual features~ \cite{chatz2022,jonmohamadi2020extraction}. 
	
	CTD has been also considered in an FL setting (e.g., \cite{kim2017federated, ma2021communication1}) in view of its intrinsic ability to extract common and distinct features, and effectively prevent privacy leakage. %Leveraging the potential of CTD, it is posited that the framework of coupled tensors can holistically elucidate and formulate such instances of FL. 
	Moreover, in such a distributed TD context, it enables efficient processing of large-scale data as demonstrated in numerous related studies, e.g., ~\cite{kang2012gigatensor,choi2014dfacto,beutel2014flexifact,zhe2016distributed,shin2016fully,aggour2018accelerating}. %Therefore, FL based on CTD is the subject area of this paper. 
	
	\subsection{Related Work}
	
Existing related work has predominantly concentrated on optimizing communication costs and ensuring privacy protection. Thus, Kim \emph{et al.}~\cite{kim2017federated} demonstrated CPD of multiple hospital data without sharing patient-specific information. To further enhance privacy preservation, Ma \emph{et al.}~\cite{ma2019privacy} implemented federated TD under the constraints of centralized differential privacy. In addition, Ma \emph{et al.}~\cite{ma2021communication1} proposed a communication-efficient federated setting for CPD by leveraging periodic communication strategies. \cite{ma2021communication} considered ring and star FL network topologies, and implemented a decentralized CPD using gradient compression and strategies of event-driven communication. Wang \emph{et al.}~\cite{wang2022tensor} introduced personalized FL based on TD that reduces communication overhead by only transmitting low-dimensional factor matrices. Li \emph{et al.}~\cite{li2022personalized} applied federated tensor decomposition to a heterogeneous distributed Internet Quality-of-service (QoS) prediction of the Things(IoT) service. However, all these methods are based on the CPD model~\cite{sdfhpf17}, which, despite its simplicity and mild uniqueness properties, may suffer from low accuracy and computational inefficiency, especially in the commonly large-scale settings of FL. 
% Furthermore, for the case where all modes of different tensors are coupled, Feng \emph{et al.}~\cite{feng2018privacy} implemented a privacy-preserving Tucker decomposition in a federated cloud environment. Gao \emph{et al.}~\cite{gao2021federated} proposed a joint high-order  
%  orthogonal iteration in a federated environment based on the Tucker model. In this paper, we focus on the case where only the first mode of the tensor is uncoupled.
	
	\subsection{Why Tensor-Train Decomposition?}
	
	Tensor-train decomposition (TTD)~\cite{oseledets2011tensor} can be regarded as a hierarchical tensor network structure~\cite{wang2019principal} that represents an $N$-way tensor as the contraction of a series/train of $N$ core tensors that are of order~3 except for those at the ends of the train that are of order~2. The tensor train (TT) model offers several distinct advantages. The feature extraction capabilities are improved~\cite{gao2023multi}. Compared with CPD, TT enjoys increased accuracy stability~\cite{ZNIYED2020304}. Moreover, the estimation of the ranks of the TT-cores is easier, compared with the problem of tensor rank estimation in CPD~\cite{8989951}. In comparison with the Tucker model, TTD achieves increased representation compactness~\cite{ZNIYED2020304}. Most importantly, as it only involves low-order $(\leq 3)$ tensors, TTD manages to mitigate the curse of the dimensionality problem. %Additionally, it facilitates adaptive selection of the appropriate TT rank in advance~\cite{oseledets2011tensor,wang2020adtt,gati2020differentially,bhattarai2020distributed}. 
	Lastly, expressing the tensors in the TT format facilitates a number of numerical operations~\cite{oseledets2011tensor}, including the tensor contraction product, which in turn allows the practical implementation of large tensor networks as demonstrated in~\cite{TCP}.
	
	\subsection{Our Contributions}
	
	In this paper, we investigate an alternative approach to extracting common features from coupled data in an FL network, which is based on the TT model instead and is specifically designed to ensure the privacy preservation of information across different network nodes. Compared with other CPD-based FL methods, the proposed method reduces computation time and communication rounds significantly, without compromising accuracy. As demonstrated in a classification experiment with medical data, CTT performs comparably with its centralized counterpart as far as classification accuracy is concerned. The key contributions of this work can be summarized as follows:
	
	\begin{itemize}
		\item We introduce a coupled TTD model, called \emph{coupled tensor train (CTT)}, which can couple an arbitrary number of tensors of any order. As it is depicted in Fig.~\ref{fig:CTT}, the proposed framework aims to extract global features across different nodes by sharing the cores for the \emph{feature} modes across the network while keeping the cores for the \emph{personal} modes private to their nodes. 
		\item Based on this framework, we develop a privacy-preserving distributed CTT method, tailored to meet the requirements of FL. Concretely, at each node of the FL network, the cores for the feature and personal modes of the local tensor are approximated. Then, the cores for the shared modes are aggregated to help extract the global features, which are re-distributed to all nodes, thus implementing FL. 
		\item The CTT approach is instantiated for the master-slave and decentralized network structures.  
		\item We compare the accuracy, computational complexity, and communication efficiency of the proposed method with existing methods with both synthetic and real datasets. Moreover, we experimentally investigate the algorithm parameter settings, and evaluate the impact of missing data, network topology, size and density on the performance of CTT. Notably, the loss in feature extraction performance from adopting CTT over its centralized counterpart is demonstrated in a real data classification task to be negligible. 
	\end{itemize}
	
	\subsection{Organization of the Paper}
	
	The rest of this paper is organized as follows. The notations adopted are described below in this section. Section~\ref{sec:prel} briefly defines tensor-related quantities and reviews the TT model. The problem of CTT is stated in Section~\ref{sec:problem}. Section~\ref{sec:method} develops the proposed method along with its master-slave and de-centralized versions. The computational and communication costs as well as the privacy preservation ability of the proposed approach are analyzed in Section~\ref{sec:analysis}. Section~\ref{sec:experiments} reports and discusses the experimental results, with both synthetic and real data. We conclude with a summary of our findings in Section~\ref{sec:conclusion}.
	
	\subsection{Notations}
	
	Vectors, matrices, and higher-order tensors are denoted by boldface lowercase letters $\mathbf{x}$ and boldface capital letters $\mathbf{X}$, and calligraphic letters $\mathcal{X}$, respectively. The $n$-unfolding of a tensor $\mathcal{X}$ is written as $\mathbf{X}_{\langle n\rangle}$. %The outer product is denoted by $\circ$. 
	The symbol $\boxtimes_{L}$ is used to denote the contraction of two tensors along their common $L$ modes. %$\|\cdot\|$ is the (vector) Euclidean norm while 
	$\|\cdot\|_{\F}$ stands for the Frobenius norm. 
	The identity matrix and the column vector of all ones are respectively 
	denoted by~$\mathbf{I}$ and~$\mathbf{1}$. 
	The diagonal matrix with the vector $\mathbf{a}$ on its main diagonal is written as $\mathrm{diag}(\mathbf{a})$. Transposition is denoted by $^{\T}$. 
	%$\mathbb{E}(\cdot)$ is the expectation operator. 
	The field of real numbers is denoted by $\mathbb{R}$. 
	
	\section{Preliminaries}
	\label{sec:prel}
	
	\begin{definition}[Tensor~\cite{sdfhpf17}]
		A multi-indexed array $\mathcal{X}\in\mathbb{R}^{I_1\times I_2\times \cdots\times I_N}$ with entries $\mathcal{X}(i_1,i_2,\ldots,i_N)$, $i_n=1,2,\ldots,I_n$, $n=1,2,\ldots,N$ is referred to as an $N$th-order or $N$-way tensor with $N$ modes of sizes $I_1,I_2,\ldots,I_N$. Thus, scalars, vectors, and matrices are tensors of order~0, 1, and~2, respectively. By higher-order tensor, we will mean a tensor of order $N>2$.
		\label{def:tensor}
	\end{definition}
	
	\begin{definition}[$n$-unfolding~\cite{sdfhpf17}]
		The $n$-unfolding of a tensor $\mathcal{X}\in\mathbb{R}^{I_1\times I_2\times \cdots\times I_N}$ is the $I_n\times \prod_{i\neq n}I_i$ matrix $\mathbf{X}_{\langle n \rangle}$ that contains the mode-$n$ vectors of $\mathcal{X}$ as its columns. The mode-$n$ vectors of $\mathcal{X}$ is in any order, provided it is consistent.
		\label{def:unfolding}
	\end{definition}
	
	\begin{definition}[Tensor contraction product~\cite{sdfhpf17}]
		Given the tensors $\mathcal{X}\in\mathbb{R}^{I_1\times\cdots\times I_N\times J_1\times\cdots\times J_L}$ and $\mathcal{Y}\in\mathbb{R}^{J_1\times\cdots\times J_L \times K_1\times\cdots\times K_M}$, their contraction along their common $L$ modes yields a new tensor, $\mathcal{S}\triangleq \mathcal{X}\boxtimes_{L}\mathcal{Y}\in\mathbb{R}^{I_1\times\cdots\times I_N\times K_1\times\cdots\times K_M}$, defined as
		\begin{equation}
			\begin{split}
				\mathcal{S}(i_1,...,i_N, k_1,...,k_M)=\sum_{j_1,...,j_L}&\mathcal{X}(i_1,...,i_N,j_1,...,j_L)\\
				&\times\mathcal{Y}(j_1,...,j_L,k_1,...,k_M).
			\end{split}
		\end{equation}
		This operation is visualized in Fig.~\ref{fig:tensor contraction}, using tensor network notation~\cite{TCP}. Clearly, $\boxtimes_{L}$ reduces to the well-known matrix product when $L=M=N=1$. For the sake of simplicity, $\boxtimes_{1}$ will be simplified to $\boxtimes$ in the following.
		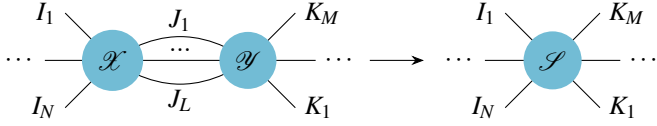
\begin{figure}
			\centering
			\begin{tikzpicture}[scale=0.9]
				\definecolor{Color1}{RGB}{114,188,213}
				\draw[] (0,0)  parabola[ parabola height=10 ](2,0);
				\node (J1) at (1,17pt) {$J_1$};
				\draw[] (0,0)  parabola[parabola height=-10](2,0);
				\node (JL) at (1,-17pt) {$J_L$};
				\draw[] (-1,0)node [left]{$\cdots$}  --node [above]{$...$}(3,0)node [right]{$\cdots$};
				\draw (-0.7,0.7)node [left]{$I_1$} --(0,0)
				[insert path={node[fill=Color1,circle]{$\mathcal{X}$}}]
				--(-0.7,-0.7)node [left]{$I_N$};
				\draw (2.7,0.7)node [right]{$K_M$} --(2,0)
				[insert path={node[fill=Color1,circle]{$\mathcal{Y}$}}]
				--(2.7,-0.7)node [right]{$K_1$};
				\draw [arrows = {-Stealth[reversed, reversed]}] (3.8,0) -- (4.6,0);
				\draw (7.2,0.7)node [right]{$K_M$} --(6.5,0)--(7.2,-0.7)node [right]{$K_1$};
				\draw[] (5.5,0)node [left]{$\cdots$}  --node [above]{$...$}(7.5,0)node [right]{$\cdots$};
				\draw (5.8,0.7)node [left]{$I_1$} --(6.5,0)
				[insert path={node[fill=Color1,circle]{$\mathcal{S}$}}]
				--(5.8,-0.7)node [left]{$I_N$};
			\end{tikzpicture}\\
			\caption{Tensor network illustration of the contraction of two tensors of size $I_1\times\cdots\times I_N\times J_1\times\cdots\times J_L$ and $J_1\times\cdots\times J_L \times K_1\times\cdots\times K_M$.}
			\label{fig:tensor contraction}
		\end{figure}
		\label{def:contraction}
	\end{definition}
	
	\begin{definition}[Tensor Train Decomposition (TTD)~\cite{oseledets2011tensor}]
		As it shows in Fig.~\ref{fig:TT}, the tensor-train decomposition (TTD) of an $N$th-order tensor $\mathcal{X}\in\mathbb{R}^{I_1\times I_2\times\cdots\times I_N}$ writes its entries as
		\begin{equation}
			\mathcal{X}(i_1,i_2,\ldots,i_N)=\mathbf{G}^{(1)}(i_1)\mathbf{G}^{(2)}(i_2)\cdots\mathbf{G}^{(N)}(i_N),
		\end{equation}
		where, for each given $i_n$, $\mathbf{G}^{(n)}(i_n)\in\mathbb{R}^{R_{n-1}\times R_{n}}$, $n = 1,2,\ldots,N$, and $R_0=R_N=1$. These matrices can be regarded as the lateral slices of 3rd-order tensors $\mathcal{G}_{n}\in\mathbb{R}^{R_{n-1}\times I_n\times R_n}$, known as TT-cores, where $\mathcal{G}_1\in\mathbb{R}^{I_1\times R_1}$ and $\mathcal{G}_N\in\mathbb{R}^{R_{N-1}\times I_N}$ are matrices. Using tensor contraction notation, the TTD can also be written as 
		\begin{equation}
			\mathcal{X}=\mathcal{G}_1\boxtimes \mathcal{G}_2\boxtimes \cdots \boxtimes \mathcal{G}_{N-1}\boxtimes \mathcal{G}_N.
		\end{equation}
		In analogy with the multilinear rank for the Tucker decomposition, the TT-rank of a tensor $\mathcal{X}\in\mathbb{R}^{I_1\times\cdots\times I_N}$ is defined as
		\begin{equation}
			\text{rank}_{\text{TT}}(\mathcal{X})=\left[R_0,R_1,\ldots,R_{N}\right],
		\end{equation}
		where $R_n\leq\text{rank}(\mathcal{X}_{\langle n\rangle})$ for $n = 1,\ldots, N$~\cite[Theorem~2.1]{oseledets2011tensor}.
		\begin{figure}[!ht]
			\centering
			\begin{tikzpicture}[scale=0.82]
				\tikz [every edge quotes/.style={fill=white,font=\small}];
				\definecolor{Color1}{RGB}{114,188,213}
				%-----------------------------------------------------------
				\draw[] (-2.7,0.8)node [right]{$\cdots$} --(-3.5,0)--(-2.7,-0.8)node [right]{$\cdots$};
				\draw[] (-2.5,0)node [right]{$I_n$} --(-3.5,0)--(-4.5,0)node [left]{$I_1$};
				\draw[] (-4.3,0.8)node [left]{$I_N$} --(-3.5,0) --(-4.3,-0.8)node [left]{$I_2$};
				\draw[fill=Color1] (-3.5,0) node{$\mathcal{X}$} circle (0.5cm) ;

				\draw[] (-1.6,0) node[]{$\approx$};
				\draw[] (-1.15,0) --node [above]{$R_0$} (-0.42,0);
				\draw[fill=Color1] (0, 0) node{$\mathcal{G}_1$} circle (0.42cm) ;
				\draw[] (0,-0.42) -- node [right]{$I_1$} (0,-1) ;
				\draw[] (0.42,0) -- node [above]{$R_1$}(1.08,0);
				\draw[fill=Color1] (1.5, 0) node{$\mathcal{G}_2$} circle (0.42cm);
				\draw[] (1.5,-0.42) --node [right]{$I_2$} (1.5,-1);
				\draw[] (1.92,0) -- node [above]{$R_2$} (2.65,0);
				
				\draw[] (3,0) node[]{$\cdots$};
				
				\draw[] (3.35,0) -- node [above]{$R_{N-1}$} (4.08,0);
				\draw[fill=Color1] (4.5,0) node{$\mathcal{G}_N$} circle (0.42cm);
				\draw[] (4.5,-0.42) --node [right]{$I_N$} (4.5,-1);
				\draw[] (4.92,0) -- node [above]{$R_{N}$}(5.7,0);
				
				% \draw[] (6,0) node[]{$\cdots$};
				
				% \draw[] (6.35,0) -- node [above]{$R_N$}(7.15,0);
				% \draw[fill=Color1] (7.5, 0) circle (0. 35cm);
				% \draw[] (7.5,-0.35) --node [right]{$I_N$} (7.5,-1);
				% \draw[] (7.85,0) --node [above]{$R_{N+1}$} (8.65,0);
				
				%---------------------------------------------------
			\end{tikzpicture}
			\caption{Tensor network illustration of the TT decomposition for an $I_1\times I_2\times\cdots\times I_N$ tensor.}
			\label{fig:TT}
		\end{figure}
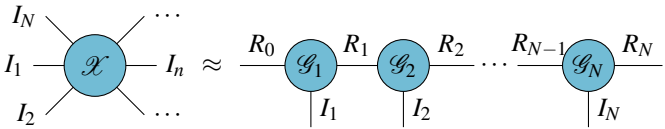
		\label{def:TT}
	\end{definition}
	
	\subsection{Tensor-Train Singular Value Decomposition (TT-SVD)}
	
	Tensor-Train Singular Value Decomposition (TT-SVD) is the best-known procedure for fitting a TTD to a given tensor $\mathcal{X}\in\mathbb{R}^{I_1\times I_2\times\cdots\times I_N}$ in the least-squares sense. In addition, as its name suggests, it comprises a series of truncated singular value decompositions (SVD) of the successive unfoldings, which also reveal the upper bounds of the TT-rank. Since this will be our TTD computing tool in this work, it is briefly summarized below. For more details on this method, the reader is referred to~\cite{oseledets2011tensor}.
	
	To attain a prescribed accuracy, $\epsilon$, namely $\frac{\|\mathcal{X}-\mathcal{\hat{X}}\|_{\F}}{\|\mathcal{X}\|_{\F}}\leq \epsilon$, where $\mathcal{X}$ is the TT tensor, the truncation parameter for the successive truncated SVDs must be computed as~\cite{oseledets2011tensor}
	\begin{equation}
		\delta=\frac{\epsilon}{\sqrt{N-1}}\|\mathcal{X}\|_{\F}.
	\end{equation}
	Then, the singular values at each SVD must be truncated at $\delta$ and the TT-rank estimates will be the corresponding $\delta$-ranks~\cite{oseledets2011tensor}. The procedure starts with the 1-unfolding, 
	$\mathbf{C}\triangleq \mathbf{C}_{\langle 1\rangle}\in\mathbb{R}^{I_1\times I_2\cdots I_N}$, of the tensor $\mathcal{C}\triangleq \mathcal{X}$ and its $\delta$-truncated SVD:
	\begin{equation}
		\mathbf{C}=\mathbf{U}\mathbf{S}\mathbf{V}^{\T}+\mathbf{E},
		\label{eq:tSVD}
	\end{equation}
	where $\|\mathbf{E}\|_{\F}\leq\delta$ and $R_1=\text{rank}_{\delta}(\mathbf{C})$. 
	$\mathbf{U}\in\mathbb{R}^{I_1\times R_1}$ yields the first TT-core, $\mathcal{G}_1$. In order to compute the second TT-core, the matrix $\mathbf{S}\mathbf{V}^{\T}\in\mathbb{R}^{R_1\times I_2\cdots I_N}$ is reshaped into a tensor, call it again $\mathcal{C}\in\mathbb{R}^{R_1I_2\times I_3\times\cdots\times I_N}$ of order $N-1$, and its 1-unfolding matrix, $\mathbf{C}\triangleq \mathbf{C}_{\langle 1\rangle}\in\mathbb{R}^{R_1I_2\times I_3\cdots I_N}$ undergoes the same processing as in~\eqref{eq:tSVD}, where $R_2\triangleq \text{rank}_{\delta}(\mathbf{C})$.
	Then, TT-SVD reshapes the new $\mathbf{U}\in\mathbb{R}^{R_1I_2\times R_2}$ into $\mathcal{G}_2$ of size $R_1\times I_2\times R_2$ and continues with the new $\mathbf{S}\mathbf{V}^{\T}\in\mathbb{R}^{R_2\times I_3\cdots I_N}$ until all TT-cores and TT-ranks are obtained. This procedure is summarized in Alg.~\ref{alg:TT_SVD} using Matlab\textcopyright\ notation and will henceforth be referred to also as TT-SVD($\epsilon$). 
	\begin{algorithm}
		\caption{TT-SVD($\epsilon$)}
		\hspace*{0.02in} {\bf Input:} Tensor $\mathcal{X}\in\mathbb{R}^{I_{1}\times I_{2}\times \cdots \times I_{N}}$, prescribed relative accuracy $\epsilon$.\\
		\hspace*{0.02in} {\bf Output:} TT-cores $\mathcal{G}_{1},\mathcal{G}_{2},\ldots,\mathcal{G}_{N}$.
		
		\begin{algorithmic}[1]
			
			\State 	\textbf{Initialization:} Compute truncation parameter $\delta=\frac{\epsilon}{\sqrt{N-1}}\|\mathcal{X}\|_{\F}$
			\State 	\textbf{Temporary tensor:} $\mathcal{C}=\mathcal{X}$, $R_0=1$
			\For{$n=1$ to $N-1$}
			\State  $\mathbf{C} \triangleq \text{\tt reshape}\left(\mathcal{C},R_{n-1}I_{n},[\;]\right)$ 
			\State Compute $\delta$-truncated SVD: $\mathbf{C}=\mathbf{U}\mathbf{S}\mathbf{V}^{\T}+\mathbf{E},\|\mathbf{E}\|_{\F}\leq\delta, R_{n}=\text{rank}_{\delta}(\mathbf{C})$
			\State New core: $\mathcal{G}_{n} \triangleq \text{\tt reshape}(\mathbf{U},\left[R_{n-1},I_{n},R_n\right])$
			\State $\mathcal{C} \triangleq \text{\tt reshape}\left(\mathbf{S}\mathbf{V}^{\T},\left[R_{n},I_{n+1},I_{n+2},\ldots,I_{N}\right]\right)$ 
			\EndFor
			\State \textbf{end for}
			\State $\mathcal{G}_{N}=\mathcal{C}$
		\end{algorithmic}
		\label{alg:TT_SVD}
	\end{algorithm}
	
	\section{Problem Statement and Solution}
	\label{sec:problem}
	
	Consider a network of $K$ nodes, and $k = 1, 2, \ldots, K$, referred to as clients, each stores its own data in the form of an $N$th-order tensor, $\mathcal{X}^k\in\mathbb{R}^{I_1^k\times I_2\times \cdots\times I_N}$. Thus, data share a number of $N-1$ common dimensions/modes/ways. For example, as in~\cite{kim2017federated}, clients may represent different hospitals keeping EHRs with ``patient", ``medication", and ``diagnosis" modes stored in 3rd-order tensors. These should share information about medication and diagnosis in order to help extract useful clinical features while at the same time making sure that patient-sensitive information, found in the personal mode, is kept private to each hospital.
	
	This is clearly a CTD problem, with the additional requirement of privacy preservation. Moreover, due to storage constraints, CTD must be performed in a distributed manner, not by centralizing all client data to a server. Thus, as in FL, local model computations must be initially performed on the client sides, followed by the aggregation of feature modes on the server side, which allows for the extraction of useful common information when the aggregated tensor is further decomposed back in the clients. 
	
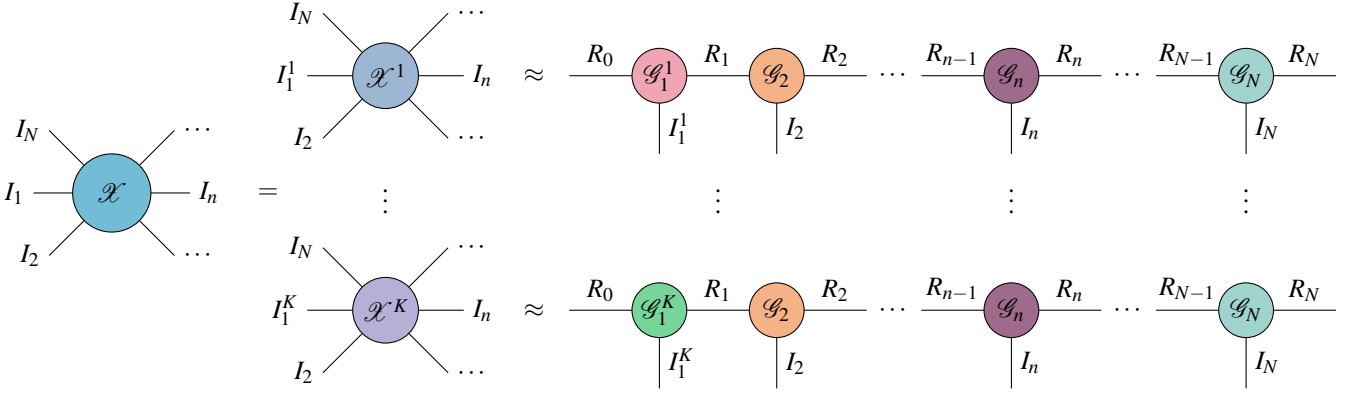
\begin{figure*}
		\centering
		
		\begin{tikzpicture}[scale=1.04]
			\tikz [every edge quotes/.style={fill=white,font=\small}];
			\definecolor{Color1}{RGB}{114,188,213}
			\definecolor{Color2}{RGB}{239,165,180}
			\definecolor{Color3}{RGB}{119,213,153}
			\definecolor{Color4}{RGB}{159,107,141}
			\definecolor{Color5}{RGB}{245,178,135}
			\definecolor{Color6}{RGB}{161,211,206}
			\definecolor{Color7}{RGB}{157,183,210}
			\definecolor{Color8}{RGB}{181,176,214}
			\draw[] (-6.2,-0.7)node [right]{$\cdots$} --(-7,-1.5)--(-6.2,-2.3)node [right]{$\cdots$};
			\draw[] (-6,-1.5)node [right]{$I_n$} --(-7,-1.5)--(-8,-1.5)node [left]{$I_1$};
			\draw (-7.8,-0.7)node [left]{$I_N$} --(-7,-1.5)--(-7.8,-2.3)node [left]{$I_2$};
			\draw[fill=Color1] (-7,-1.5) node{$\mathcal{X}$} circle (0.5cm) ;
			\draw[] (-5,-1.5) node[]{$=$};
			%-----------------------------------------------------------
			\draw[] (-2.7,0.8)node [right]{$\cdots$} --(-3.5,0)--(-2.7,-0.8)node [right]{$\cdots$};
			\draw[] (-2.5,0)node [right]{$I_n$} --(-3.5,0)--(-4.5,0)node [left]{$I_1^1$};
			\draw[] (-4.3,0.8)node [left]{$I_N$} --(-3.5,0) --(-4.3,-0.8)node [left]{$I_2$};
			\draw[fill=Color7] (-3.5,0) node{$\mathcal{X}^1$} circle (0.42cm) ;

			\draw[] (-1.6,0) node[]{$\approx$};
			\draw[] (-1.15,0) --node [above]{$R_0$} (-0.35,0);
			\draw[fill=Color2] (0, 0) node{$\mathcal{G}_1^1$}circle (0.35cm) ;
			\draw[] (0,-0.35) -- node [right]{$I_1^1$} (0,-1) ;
			\draw[] (0.35,0) -- node [above]{$R_1$}(1.15,0);
			\draw[fill=Color5] (1.5, 0) node{$\mathcal{G}_2$}circle (0.35cm);
			\draw[] (1.5,-0.35) --node [right]{$I_2$} (1.5,-1);
			\draw[] (1.85,0) -- node [above]{$R_2$} (2.65,0);
			
			\draw[] (3,0) node[]{$\cdots$};
			
			\draw[] (3.35,0) -- node [above]{$R_{n-1}$} (4.15,0);
			\draw[fill=Color4] (4.5, 0)node{$\mathcal{G}_n$} circle (0.35cm);
			\draw[] (4.5,-0.35) --node [right]{$I_n$} (4.5,-1);
			\draw[] (4.85,0) -- node [above]{$R_{n}$}(5.65,0);
			
			\draw[] (6,0) node[]{$\cdots$};
			
			\draw[] (6.35,0) -- node [above]{$R_{N-1}$}(7.15,0);
			\draw[fill=Color6] (7.5, 0) node{$\mathcal{G}_N$}circle (0. 35cm);
			\draw[] (7.5,-0.35) --node [right]{$I_N$} (7.5,-1);
			\draw[] (7.85,0) --node [above]{$R_{N}$} (8.65,0);
			
			%---------------------------------------------------
			\draw[] (-3.5,-1.5) node[]{$\vdots$};
			\draw[] (0.75,-1.5) node[]{$\vdots$};
			\draw[] (4.5,-1.5) node[]{$\vdots$};
			\draw[] (7.5,-1.5) node[]{$\vdots$};
			%---------------------------------------------------
			
			\draw[] (-2.7,-2.2)node [right]{$\cdots$} --(-3.5,-3)--(-2.7,-3.8)node [right]{$\cdots$};
			\draw[] (-2.5,-3)node [right]{$I_n$} --(-3.5,-3)--(-4.5,-3)node [left]{$I_1^K$};
			\draw (-4.3,-2.2)node [left]{$I_N$} --(-3.5,-3)--(-4.3,-3.8)node [left]{$I_2$};
			\draw[fill=Color8] (-3.5,-3) node{$\mathcal{X}^K$} circle (0.42cm) ;

			\draw[] (-1.6,-3) node[]{$\approx$};
			\draw[] (-1.15,-3) --node [above]{$R_0$} (-0.35,-3);
			\draw[fill=Color3] (0, -3)  node{$\mathcal{G}_1^K$}circle (0.35cm) ;
			\draw[] (0,-3.35) -- node [right]{$I_1^K$} (0,-4) ;
			\draw[] (0.35,-3) -- node [above]{$R_1$}(1.15,-3);
			\draw[fill=Color5] (1.5, -3)  node{$\mathcal{G}_2$}circle (0.35cm);
			\draw[] (1.5,-3.35) --node [right]{$I_2$} (1.5,-4);
			\draw[] (1.85,-3) -- node [above]{$R_2$} (2.65,-3);
			
			\draw[] (3,-3) node[]{$\cdots$};
			
			\draw[] (3.35,-3) -- node [above]{$R_{n-1}$} (4.15,-3);
			\draw[fill=Color4] (4.5, -3)  node{$\mathcal{G}_n$}circle (0.35cm);
			\draw[] (4.5,-3.35) --node [right]{$I_n$} (4.5,-4);
			\draw[] (4.85,-3) -- node [above]{$R_{n}$}(5.65,-3);
			
			\draw[] (6,-3) node[]{$\cdots$};
			
			\draw[] (6.35,-3) -- node [above]{$R_{N-1}$}(7.15,-3);
			\draw[fill=Color6] (7.5, -3)  node{$\mathcal{G}_N$}circle (0. 35cm);
			\draw[] (7.5,-3.35) --node [right]{$I_N$} (7.5,-4);
			\draw[] (7.85,-3) --node [above]{$R_{N}$} (8.65,-3);
		\end{tikzpicture}
		\caption{The network architecture of CTT. $\mathcal{X}^k$: the tensor stored in the $k$th client. $\mathcal{G}_1^k$: the TT-core for the personal mode of the $k$th client. $\mathcal{G}_n$: the TT-core for the $n$th global feature mode, $n=2,\ldots,N$.}
		\label{fig:CTT}
	\end{figure*}

	In this work, we propose to follow a coupled TTD approach, illustrated in Fig.~\ref{fig:CTT}, where tensors from different clients in modes $n=2,\ldots,N$ are coupled. 
	Each client tensor $\mathcal{X}^k$ is decomposed into core tensors, where $\mathcal{G}_1^k$ represents the personal mode, while $\mathcal{G}_n$, $n\geq2$ correspond to the feature modes and are shared among all tensors. For the sake of simplicity, we assume that all $R_1^k$ ranks coincide with $R_1$. If this is not the case, one would have to somehow \emph{align} the $K$ personal modes. An example solution to this problem is given in~\cite{kim2017federated}. We will neglect this issue here and return to it in future work. For now, we may assume that, based on \emph{a-priori} information on the personal mode, one can set an appropriate common value for the $R_1^k$s.
	
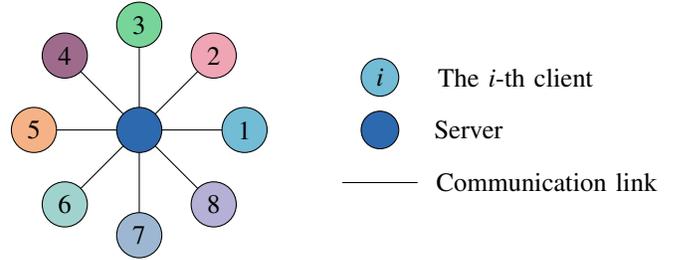
\begin{figure}
		\centering
		\begin{tikzpicture}
			\definecolor{Color1}{RGB}{114,188,213}
			\definecolor{Color2}{RGB}{239,165,180}
			\definecolor{Color3}{RGB}{119,213,153}
			\definecolor{Color4}{RGB}{159,107,141}
			\definecolor{Color5}{RGB}{245,178,135}
			\definecolor{Color6}{RGB}{161,211,206}
			\definecolor{Color7}{RGB}{157,183,210}
			\definecolor{Color8}{RGB}{181,176,214}
			\definecolor{Color9}{RGB}{45,105,174}
			\draw[] (0, 0) -- (xy polar cs:angle=0,radius=1.4);
			\draw[] (0, 0) -- (xy polar cs:angle=45,radius=1.4);
			\draw[] (0, 0) -- (xy polar cs:angle=90,radius=1.4);
			\draw[] (0, 0) -- (xy polar cs:angle=135,radius=1.4);
			\draw[] (0, 0) -- (xy polar cs:angle=180,radius=1.4);
			\draw[] (0, 0) -- (xy polar cs:angle=225,radius=1.4);
			\draw[] (0, 0) -- (xy polar cs:angle=270,radius=1.4);
			\draw[] (0, 0) -- (xy polar cs:angle=315,radius=1.4);
			\draw[fill=Color9] (0, 0) circle (0.3cm) ;
			\draw[fill=Color1] (xy polar cs:angle=0,radius=1.4) circle (0.3cm) node{1};
			\draw[fill=Color2] (xy polar cs:angle=45,radius=1.4) circle (0.3cm) node{2};
			\draw[fill=Color3] (xy polar cs:angle=90,radius=1.4) circle (0.3cm) node{3};
			\draw[fill=Color4] (xy polar cs:angle=135,radius=1.4) circle (0.3cm) node{4} ;
			\draw[fill=Color5] (xy polar cs:angle=180,radius=1.4) circle (0.3cm) node{5};
			\draw[fill=Color6] (xy polar cs:angle=225,radius=1.4) circle (0.3cm) node{6};
			\draw[fill=Color7] (xy polar cs:angle=270,radius=1.4) circle (0.3cm)node{7};
			\draw[fill=Color8] (xy polar cs:angle=315,radius=1.4) circle (0.3cm) node{8};
			\draw[fill=Color1] (3.2,0.7) circle (0.25cm) node [label=right:$\quad$ The $i$-th client] {$i$};
			\draw[fill=Color9] (3.2,0) circle (0.25cm) node [label=right:$\quad$ Server] {};
			\draw[] (2.7,-0.7)-- (3.7,-0.7)  node [label=right: Communication link] {};
		\end{tikzpicture}
		\caption{Illustration of a master-slave network structure.}
		\label{fig:Master-slave}
	\end{figure}

\begin{figure}
		\centering
		\begin{tikzpicture}
			\definecolor{Color1}{RGB}{114,188,213}
			\definecolor{Color2}{RGB}{239,165,180}
			\definecolor{Color3}{RGB}{119,213,153}
			\definecolor{Color4}{RGB}{159,107,141}
			\definecolor{Color5}{RGB}{245,178,135}
			\definecolor{Color6}{RGB}{161,211,206}
			\definecolor{Color7}{RGB}{157,183,210}
			\definecolor{Color8}{RGB}{181,176,214}
			\draw[] (0, 0) -- (-0.96, -0.24) --(-1.2, 1.2) --(-0.24, 1.12) --(0, 0) --(0.96, -1.04) --(2.24, -1.28)--(1.92, -0.24)--(2.96, -0.4) --(2.24, -1.28);
			\draw[fill=Color4] (0, 0) circle (0.3cm) node{4};
			\draw[fill=Color1] (-1.2, 1.2) circle (0.3cm) node{1};
			\draw[fill=Color3] (-0.96, -0.24) circle (0.3cm) node{3};
			\draw[fill=Color2] (-0.24, 1.12) circle (0.3cm) node{2};
			\draw[fill=Color5] (0.96, -1.04) circle (0.3cm) node{5};
			\draw[fill=Color6] (2.24, -1.28) circle (0.3cm) node{6};
			\draw[fill=Color8] (1.92, -0.24) circle (0.3cm) node{8};
			\draw[fill=Color7] (2.96, -0.4) circle (0.3cm)node{7};
		\end{tikzpicture}
		\caption{Illustration of a decentralized network structure. Nodes represent clients and the lines connecting them signify that they can communicate.}
		\label{fig:Decentralized}
	\end{figure}
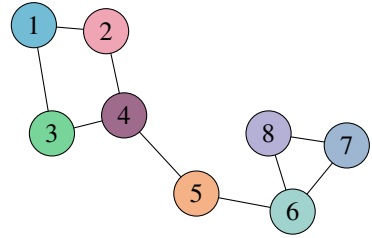

	As in, e.g., \cite{kim2017federated}, a two-way procedure is implemented, however without the iterations suggested therein: first, on the client side, the tensors are TTD-ed and the cores of the feature modes are aggregated to obtain a global feature tensor. Second, the latter is TTD-ed to yield the features that are re-distributed to the clients. In contrast to~\cite{kim2017federated} that is restricted to the master-slave network structure, we develop CTT-based FL methods implementing the above procedure for both master-slave (Fig.~\ref{fig:Master-slave}) and decentralized (Fig.~\ref{fig:Decentralized}) networks. TTD is computed with the aid of TT-SVD.
	
	\section{The Proposed Method}
	\label{sec:method}
	
	The two-step procedure briefly described previously is detailed below. 
	
	\begin{itemize} 
		\item Given the precision $\epsilon_1$, we compute the truncation parameter for each client,
		\[
		\delta_1^k=\frac{\epsilon_1}{\sqrt{N-1}}\|\mathcal{X}^k\|_{\F}.
		\]
		Then we perform the first truncated SVD in the TT-SVD sequence:
		\begin{equation}
			\label{eq:svd}
			\mathbf{X}_{\langle 1\rangle}^{k}=\mathbf{U}_1^k\mathbf{S}_1^k(\mathbf{V}_1^k)^{\T}+\mathbf{E}_1^k\equiv \mathbf{U}_1^k\mathbf{D}_1^k+\mathbf{E}_1^k, \quad \|\mathbf{E}_1^k\|_{\F}\leq \delta_1^k,
		\end{equation}
		where $\mathbf{U}_1^k\in\mathbb{R}^{I_{1}^k\times R_1}$ has orthonormal columns. The local ``personal" core then results as $\mathcal{G}_1^k=\mathbf{U}_1^k$.
		
		\item The aggregated tensor $\mathcal{W}\in \mathbb{R}^{R_1 I_2\times\cdots\times I_N}$ of the feature modes is computed from
		\begin{equation}
			\min_{\mathcal{W}}\Psi=\sum_{k=1}^{K}\|\mathcal{X}^k-\mathcal{G}_1^k\boxtimes\mathcal{W}\|_{\F}^2,
			\label{eq:W}
		\end{equation}
		which can be equivalently written as
		\[
		\min_{\mathbf{W}_{\langle 1\rangle}}\Psi=\sum_{k=1}^{K}\|\mathbf{X}^k_{\langle 1\rangle}-\mathbf{U}_1^k\mathbf{W}_{\langle 1\rangle}\|_{\F}^2.
		\] 
		This results in
		\[
		\mathbf{W}_{\langle 1\rangle}=\frac{1}{K}\sum_{k=1}^{K}(\mathbf{U}_1^k)^{\T} \mathbf{X}^k_{\langle 1\rangle},
		\]
		which, for high SVD precision, can in turn be written as
		\begin{equation}
			\mathbf{W}_{\langle 1\rangle}\approx\frac{1}{K}\sum_{k=1}^{K}\mathbf{D}_1^k.
			\label{eq:optimal W}
		\end{equation}
		We then need to learn the feature modes from $\mathbf{W}_{\langle 1\rangle}$. We will answer this in the following for the two basic network structures under consideration.
	\end{itemize}
	
	\subsubsection{Master-slave Network}
	
	The master-slave network, illustrated in Fig.~\ref{fig:Master-slave}, comprises a server, which is connected to each client via a communication link, while the clients lack direct communication links. To communicate with other clients, one must first send the message to the server, which then forwards it to the intended recipient. The server also has the capability of aggregating messages from multiple clients and issuing instructions to the clients. 
	
	Eq.~\eqref{eq:optimal W} suggests that the client $k$ transmits $\mathbf{D}_1^k$ to the server. However, this requires communicating $KR_1\prod_{n=2}^{N}I_n$ numbers in total. This communication cost can be significantly reduced, at the cost of additional computation at the clients, if TT-SVD is completed at each client node and the resulting feature mode cores, $\mathcal{G}_n^k$, $n=2,\ldots,N$, of total size $K\sum_{n=2}^{N}R_{n-1} I_n R_{n}$ are transmitted instead. 
	It is readily verified that the aggregated tensor in~\eqref{eq:W} can be found at the server as
	\begin{equation}
		\mathcal{W}=\frac{1}{K}\sum_{k=1}^{K}\mathcal{G}_2^k\boxtimes\cdots\boxtimes\mathcal{G}_N^k.
		\label{eq:aggreg}
	\end{equation}
	Based on the prescribed accuracy $\epsilon_2$ and the corresponding truncation parameter $\delta_2=\frac{\epsilon_2}{\sqrt{N-2}}\|\mathcal{W}\|_{\F}$, the server can then compute the TTD of $\mathcal{W}$ and subsequently broadcast the extracted global features $\mathcal{G}_n$, $n=2,\ldots,N$ to the clients. The procedure is summarized in Alg.~\ref{alg:CTT(master-slave mode)}.
	\begin{algorithm}
		\caption{CTT (M-s)} 
		\hspace*{0.02in} {\bf Input:} 
		$\mathcal{X}^k$, $k=1,2\ldots,K$, $\epsilon_1,\epsilon_2,R_1$\\
		\hspace*{0.02in} {\bf Output:} 
		$\mathcal{G}_1^k$, $k=1,2,\ldots,K$, $\mathcal{G}_2,\mathcal{G}_3,\ldots,\mathcal{G}_N$
		\begin{algorithmic}[1]
			\State Each client $k$ performs TT-SVD($\epsilon_1$) for the $\mathcal{X}^k$ tensor
			\State Each client $k$ sends $\mathcal{G}_2^k,\ldots,\mathcal{G}_N^k$ to the server
			\State The server performs fusion according to~(\ref{eq:aggreg})
			\State The server performs TT-SVD($\epsilon_2$) for the $\mathcal{W}$ tensor
			\State The server broadcasts $\mathcal{G}_2,\mathcal{G}_3,\ldots,\mathcal{G}_N$ to the clients  	
		\end{algorithmic}
		\label{alg:CTT(master-slave mode)}
	\end{algorithm}
	
	\subsubsection{Decentralized Network}
	
	Such a network can be seen as an undirected graph with a symmetric adjacency matrix~\cite{mao2018walk} $\mathbf{M}\in \mathbb{R}^{K\times K}$, where $m_{ij}>0$ if and only if nodes $i$ and $j$ are connected and $m_{ij}=0$ otherwise. Furthermore, the sum of each row and each column of such a matrix is equal to 1. In other words, $\mathbf{M}$ is doubly stochastic:
	\begin{eqnarray}
		\mathbf{M}\mathbf{1} & = & \mathbf{1}, \label{eq:M1=1} \\
		\mathbf{1}^{\T}\mathbf{M} & = & \mathbf{1}^{\T}, \label{eq:1M=1} \\
		\mathbf{M}^{\T} & = & \mathbf{M}. \label{eq:M=MT}
	\end{eqnarray}
	Its largest eigenvalue is unity (cf.~\eqref{eq:M1=1}) and its second largest eigenvalue, $0\leq\lambda_2<1$, is intimately connected to the network's connectivity~\cite{jiao2022communication} as will be seen in the sequel. Fig.~\ref{fig:Decentralized} illustrates a decentralized network consisting of eight nodes. A possible way of defining the adjacency matrix for a decentralized network of $K$ nodes is
	\begin{equation}
		m_{ij}=\left\{
		\begin{array}{cl}
			\dfrac{1}{K}, & j\in \mathbb{N}_i\\
			\dfrac{K-d_i}{K}, & j=i\\
			0, & \text{otherwise}
		\end{array} \right.,
		\label{eq:M}
	\end{equation}
	where $d_i$ is the degree of node $i$, and $\mathbb{N}_i$ is the set of indices of its immediate neighbors. This matrix for the network of Fig.~\ref{fig:Decentralized} has $\lambda_2=0.972$. An alternative way of defining the adjacency matrix for a fully connected network is presented in Section~\ref{sec:experiments}.
	
	\emph{Average Consensus (AC)}:
	Unlike master-slave networks, decentralized networks lack server-side control and depend on a consensus mechanism to ensure agreement on the network state among nodes~\cite{jiao2022communication}. We will make use of the \emph{average} consensus concept to compute the global features by allowing each node to communicate only with its neighbors. Let $\mathbf{Z}^k[0]$ be the initial state at node $k$. AC is achieved through an iterative procedure, $\mathbf{Z}^k[l+1]=\sum_{j=1}^{K}m_{kj}\mathbf{Z}^j[l]$, provided that $\lambda_2<1$. 
	If $\alpha_l^2=(\sum_{k=1}^{K}\|\mathbf{Z}^k[l]-\frac{1}{K}\sum_{q=1}^{K}\mathbf{Z}^q[l]\|_{\F}^2)/(\sum_{k=1}^{K}\|\mathbf{Z}^k[0]\|_{\F}^2)$ is the consensus error, that is, the difference from the AC, then $\alpha_l$ can be smaller than $\alpha$ if (see~\cite{1638541,jiao2022communication} and references therein):
	\begin{equation}
		L=\mathcal{O}\left(\frac{1}{\text{log}\left(\frac{1}{\lambda_2}\right)}\text{log}\left(\frac{1}{\alpha}\right)\right),
		\label{eq:L}
	\end{equation}
	whereby the role of $\lambda_2$ is made evident. 
	
	\emph{Decentralized CTT}: At each node, $k$, $\mathcal{G}_1^k$ and the contraction of the rest of the cores, say $\mathbf{Z}^k[0]$ in matrix form, are first computed with the aid of a truncated SVD with prescribed accuracy $\epsilon_1$. Then $L$ consensus iterations are performed with initial state $\mathbf{Z}^k[0]$ to approximate the network-wide average~\eqref{eq:aggreg}. If $L$ is large enough, we will have all $K$ $\mathbf{Z}^k[L]$ matrices approximately equal to~\eqref{eq:optimal W}. Finally, each node completes its TT-SVD with prescribed accuracy $\epsilon_2$ on the post-consensus tensor to extract the feature factors. The algorithm is outlined in Alg.~\ref{alg:CTT(Decentralized mode)}.
	\begin{algorithm}
		\caption{CTT (Dec)}
		\hspace*{0.02in} {\bf Input:} $\mathcal{X}^k$, $k=1,2,\ldots,K$, $\epsilon_1,\epsilon_2,R_1,L$\\
		\hspace*{0.02in} {\bf Output:} $\mathcal{G}_1^k,\mathcal{G}_2^k,\ldots,\mathcal{G}_N^k$
		\begin{algorithmic}[1]
			\State \textbf{For each} $k$
			\State Perform $\delta_1^k$-truncated SVD on $\mathbf{X}^k_{\langle 1 \rangle}$ to obtain $\mathcal{G}_1^k$ and $\mathbf{D}_1^k$
			\State Perform $L$ AC iterations with $\mathbf{Z}^k[0]=\mathbf{D}_1^k$
			\State Perform TT-SVD($\epsilon_2$) for $\mathbf{Z}^k[L]$ to compute $\mathcal{G}_n^k$, $n>1$
		\end{algorithmic}
		\label{alg:CTT(Decentralized mode)}
	\end{algorithm}
	
	\section{Analysis}
	\label{sec:analysis}
	
	\subsection{Computational Efficiency}
	
	Recall that the truncated SVD of an $M_1\times M_2$ matrix takes $\mathcal{O}(\max(M_1,M_2)\min(M_1,M_2)^2)$ operations when the one dimension is significantly larger than the other~\cite{li2019}. Given two tensors of size $M_1\times M_2 \times M_3$ and $M_3\times M_4 \times M_5$, computing their contraction product costs $\mathcal{O}(M_1M_2M_3M_4M_5)$. 
	
	Given the above, and making the realistic assumption that the ranks are smaller than the tensor dimensions, Line~1 in Alg.~\ref{alg:CTT(master-slave mode)} will require $\mathcal{O}(K\sum_{n=1}^{N-1}(R_{n-1}I_n)^2\prod_{i=n+1}^{N}I_i)$ operations. %$\mathcal{O}((I_1/K)^2\prod_{i=2}^{N}I_i)$. %The complexity of computing $\mathbf{D}_n^k$ is $\mathcal{O}(\sum_{n=1}^{N}((R_n^k)^2\prod_{i=n+1}^{N}I_i))$. 
	Calculating the $K$ terms in~\eqref{eq:aggreg} takes $\mathcal{O}(KR_1 \sum_{n=2}^{N-1}R_n R_{n+1}\prod_{i=2}^{n+1}I_i)$ operations and averaging the resulting $(N-1)$-way tensors costs $\mathcal{O}(KR_1\prod_{n=2}^{N}I_n)$ (Line~3). The complexity of the server performing TT-SVD on $\mathcal{W}$ (Line~4) is $\mathcal{O}(\sum_{n=2}^{N-1}(R_{n-1}I_n)^2\prod_{i=n+1}^{N}I_i)$.
	%$\mathcal{O}(\sum_{k=1}^{K}((R_{N-1}^k)^2\prod_{i=2}^{N}I_i)+\sum_{n=2}^{N}((R_{n-1}I_n)^2\prod_{i=n+1}^{N}I_i+(R_n)^2\prod_{i=n+1}^{N}I_i))$.
	
	In a decentralized network, the complexity of personal mode factor update is that of a truncated SVD, that is, for each node, $\mathcal{O}((I_1^k)^2\prod_{i=2}^{N}I_i)$ (Line~2 in Alg.~\ref{alg:CTT(Decentralized mode)}). %$\mathcal{O}((I_1/K)^2\prod_{i=2}^{N}I_i)$. 
	AC with $L$ iterations takes $\mathcal{O}(LKR_1\prod_{i=2}^N I_i)$ operations per node (Line~3), which can be reduced depending on the density of the network. The complexity of the feature mode update (Line~4) is of the order of $\mathcal{O}(\sum_{n=2}^{N-1}(R_{n-1}I_n)^2\prod_{i=n+1}^{N}I_i)$, per node. 
	%$\mathcal{O}(\sum_{n=2}^{N}(((R_{n-1}I_n)^2+(R_n)^2)\prod_{i=n+1}^{N}I_i))$.
	
	If we let all ranks and all dimensions be equal to $R$ and $I$, respectively, with the data being uniformly distributed among the $K$ nodes in their first dimension, i.e., $I_1^k=\frac{I}{K}$, and make the realistic assumption that $I\gg R$, the computational efficiencies for the master-slave and the decentralized networks turn out to be $\mathcal{O}\left(I^{N}\left[R^2\left(1+\frac{1}{K}\right)+\frac{1}{K^2}\right]\right)$ and $\mathcal{O}\left(\frac{I^{N+1}}{K^2}+R^2I^N+RLKI^{N-1}\right)$, respectively.
	Thus, the master-slave CTT scheme becomes more efficient as the number of nodes increases. The same happens for the decentralized scheme in not very large networks. 
	
	\subsection{Communication Efficiency}
	
	Communication efficiency is defined here as the total amount of numbers that need to be transmitted per node. 
	With a master-slave network structure, they are the core of the feature modes for uplink and downlink transmission. Therefore, the communication efficiency of each link is $\mathcal{O}(\sum_{n=1}^{N-1}R_{n}R_{n+1}I_{n+1})$.
	
	For the decentralized method, the variables needed to be transmitted for each communication are $\mathbf{D}_1^k\in \mathbb{R}^{R_1\times \prod_{i=2}^{N}I_i}$. Assuming that the method requires $L$ consensus steps, the communication efficiency can be calculated as $\mathcal{O}(LR_1\prod_{i=2}^{N}I_i)$.
	% According to (\ref{equ:AC_error}), it can be seen that $L=\mathcal{O}(\text{log}\frac{1}{\epsilon})$, so the communication complexity is
	% \begin{equation}
		% 	T_{\text{c,de}}=\mathcal{O}\left(\text{log}\frac{1}{\epsilon}\left(R\prod_{n=2}^{N}I_n\right)\right)
		% \end{equation}
	% where $R$ represents $R_n^k\text{ or } R_n,n=1,...,n,k=1,...K$
	
	Setting again all $R_n$ and all $I_n$ equal, the communication efficiencies of the master-slave and decentralized networks become $\mathcal{O}((N-2)R^2I)$ and $\mathcal{O}(LRI^{N-1})$, respectively. As expected, the master-slave structure prevails in terms of communication requirements. 
	
	\subsection{Privacy Analysis}
	
	We assume an \emph{honest-but-curious (HBC)} network, also known as \emph{semi-honest} adversary model~\cite{HBC}. This means that all nodes respect the constraints and rules of the task and do not tamper with the data while they can be curious about the data of neighboring nodes hoping to discover knowledge that they do not have. In this scenario, the privacy of client data can be protected since even if the server is granted exact knowledge of $\mathbf{D}_1^k$, the private nature of $\mathcal{G}_1^k$ prevents it from accessing the data of its clients. Specifically, it can be seen from~(\ref{eq:svd}) that $\mathbf{X}_{\langle 1\rangle}^{k}=\mathbf{U}_1^k\mathbf{D}_1^k+\mathbf{E}_1^k$.  Due to the unavailability of $\mathbf{U}_1^k$ and $\mathbf{E}_1^k$ at the server, the reconstruction of $\mathbf{X}_{\langle 1\rangle}^{k}$ cannot be realized. From an alternative viewpoint, it can be inferred that the data belonging to the client is indirectly encrypted through the intermediary of $\mathbf{U}_1^k$ and $\mathbf{E}_1^k$, preceding the server's execution of the feature fusion procedure.
	% In the event of an unintended leak, where the server gains access to $\mathbf{U}_1^k$ $(\mathcal{G}_1^k)$, the precise client data remains beyond the reach of the server, as it lacks access to element $\mathbf{E}_1^k$.
	
	Moreover, individual clients are unable to access the data of other clients in this setting. Consider two curious clients, say $p$ and $q$. As the feature extraction process is performed on the server side, client $p$ can only receive global features and cannot obtain the complete $\mathbf{D}_1^q$. Even if, by chance, $\mathbf{D}_1^q$ is leaked to client $p$ through the server, client $p$ would hope to deduce the data of client $q$ by utilizing~(\ref{eq:svd}) in reverse. However, since there is no direct communication link between clients $p$ and $q$, client $p$ cannot access $\mathcal{G}_1^q$ belonging to client $q$, thereby making it impossible for client $p$ to retrieve the data of client $q$. 
	
	In a decentralized network, where nodes can only communicate with their immediate neighbors, and considering the private nature of $\mathcal{G}_1^k$, a curious node certainly cannot recover complete data of other nodes from the noisy and truncated $\mathbf{D}_1^k$s.
	
	\section{Experiments}
	\label{sec:experiments}
	
	\subsection{Datasets}
	
	We conducted experiments with both synthetic and real data. For the real data, we have employed the ECG dataset\footnote{https://www.kaggle.com/datasets/devavratatripathy/ecg-dataset} and the Diabetes Health Indicators\footnote{https://www.kaggle.com/datasets/alexteboul/diabetes-health-indicators-dataset}. More details are in order:
	\begin{itemize}
		\item The synthetic data were generated by first randomly generating several sparse population feature modes matrices of standard Gaussian distribution. Then, each client randomly generated a personal mode matrix and combined the above feature modes matrices to generate a low-rank synthetic tensor through tensor operations. We consider the $200\times 30\times 30$ and $200\times 20\times 20\times 20$ tensors with a proportion of non-zero entries of 0.4 and 0.1 respectively.
		\item The ECG dataset is a publicly available dataset where each patient's ECG consists of 140 data points. We selected 1000 patient data to form a tensor of size $1000\times110\times140$, respectively for patient information, heart electrical signal, and time dimension.
		\item The Diabetes Health Indicators dataset was collected by the US Centers for Disease Control and Prevention and contains responses from 441,455 individuals from 1984 to 2015, with 330 features. We randomly selected 1000 cases, 20 physiological indicators, and 24 habitual features to form a tensor of size $1000\times20\times24$.
	\end{itemize}
	
	\subsection{Experimental Setup}
	
\begin{figure}
		\centerline{\includegraphics[width=1.0\columnwidth]{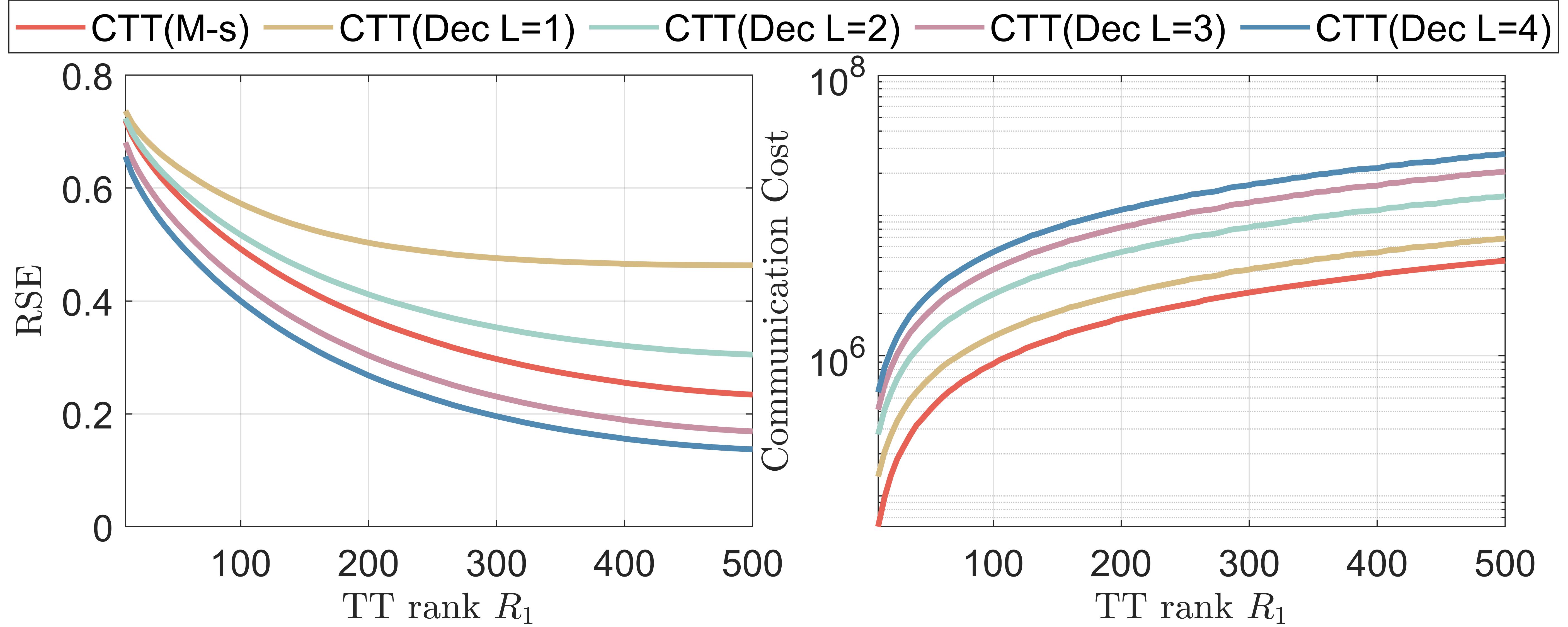}}
		\caption{Communication cost and RSE with varying $R_1$, for the ECG data. (There are $K=2$ nodes. ``M-s" and ``Dec" denote a master-slave and a decentralized fully-connected network, respectively.)}
		\label{fig:R1_ECG}
	\end{figure}

	For the implementations, we used the Tensor Toolbox Version~3.2.1~\cite{bader2021tensor}, run in Matlab~2019a on a Windows workstation with an Intel(R) Core(TM) i5-11400F~CPU@2.60~GHz processor and 16~GB of RAM. The values of the algorithm parameters were carefully selected to achieve the desired trade-off between accuracy and efficiency.
	For the relative precision parameters, we have chosen the values $\epsilon_1=0.1$ and $\epsilon_2=0.05$ after searching over the candidate values of $\epsilon_1\in\{0.05, 0.1, 0.3, 0.5,0.7\}$ and $\epsilon_2\in\{0.01, 0.03, 0.05,0.07\}$.
	How to optimize $\mathbf{M}$ or $\lambda_2$ in the decentralized network is beyond the scope of this paper. For convenience, we constructed the adjacency matrix of the fully connected network\footnote{For fully connected networks, an alternative to~\eqref{eq:M}.} by averaging the magic square matrix and its transpose to make it symmetric, and normalizing the result to satisfy eqs.~\eqref{eq:M1=1}, \eqref{eq:1M=1}. Through comparative experiments, we found that $L=3$ is a sufficiently large value for attaining a low enough consensus error. We use the relative squared error (RSE) to measure the accuracy on a given dataset $\mathcal{X}$,
	\begin{equation}
		\text{RSE}=\frac{\|\mathcal{X}-\hat{\mathcal{X}}\|_{\F}^2}{\|\mathcal{X}\|_{\F}^2},
	\end{equation}
	where $\hat{\mathcal{X}}$ is the reconstructed tensor. 
	To determine the suitable value of $R_1$ for each of the datasets, we assessed the impact of various $R_1$ values on the communication cost and RSE, as shown in Fig.~\ref{fig:R1_ECG}, Table~\ref{tab:R1_Diabetes}, and Table~\ref{tab:R1_syn}. 
	It can be seen from Fig.~\ref{fig:R1_ECG} that as $R_1$ increases, the RSE decreases while the communication cost increases. To strike a balance between communication efficiency and the capturing of site-specific features, we have seen that when the number of nodes is 4, $R_1$ values of 50, 100, and~20 are appropriate for the Diabetes, ECG, and synthetic datasets, respectively.
	\begin{table*}
		\begin{center}
			\caption{RSE, run-time, and communication cost of CTT with varying $R_1$ and $L$ ($K=4$, Diabetes data).}\label{tab:R1_Diabetes}
			{\begin{tabular}{lcccccccc}
					\toprule
					\specialrule{0pt}{2pt}{2pt}
					\multirow{2}{*}{Metric} &\multirow{2}{*}{Algorithms} & \multicolumn{7}{c}{TT rank $R_1$} \\ 
					\cline{3-9} 
					\specialrule{0pt}{2pt}{2pt}
					\multicolumn{2}{c}{}  &15  & 25    &35 &45 & 50 &60 &100  \\ 
					\midrule
					\multirow{5}{*}{RSE}  
					& CTT(M-s) &0.2518  & 0.2369    &0.2308 &0.2266 & 0.2250 &0.2228 &0.2192  \\
					& CTT(Dec L=1) & 0.3426  & 0.3448   & 0.3480  & 0.3494  &0.3498 &0.3504 &0.3521     \\
					& CTT(Dec L=2) & 0.2617  & 0.2485 & 0.2436  & 0.2400   &0.2386 &0.2367 &0.2335    \\
					& CTT(Dec L=3) &0.2523  & 0.2377    &0.2318 &0.2277 & 0.2261 &0.2239 &0.2204  \\
					& CTT(Dec L=4) &0.2519  & 0.2371 & 0.2310  & 0.2268  &0.2252 &0.2230 &0.2194     \\
					\midrule
					\multirow{5}{*}{Communication cost} 
					& CTT(M-s) &2.92e+03  & 6.29e+03    &8.69e+03 &1.11e+04 & 1.23e+04 &1.47e+04 &2.63e+04  \\
					& CTT(Dec L=1)& 7.20e+03 & 1.20e+04 & 1.68e+04  & 2.16e+04  &2.40e+04 &2.88e+04 &4.80e+04     \\
					& CTT(Dec L=2) & 1.44e+04 &2.40e+04 & 3.36e+04  &4.32e+04   &4.80e+04 &5.70e+04 &9.6e+04    \\
					& CTT(Dec L=3)      &2.16e+04  & 3.60e+04    &5.04e+04 &6.48e+04 & 7.18e+04 &8.64e+04 &1.44e+05  \\
					& CTT(Dec L=4) &2.88e+04       &4.80e+04   &6.72e+04   &8.64e+04 &9.60e+04 &1.16e+05 &1.92e+05 \\
					\midrule
					\multirow{5}{*}{CPU time (seconds)} 
					& CTT(M-s)&0.0328    &0.0379   &0.0373    &0.0391    &0.0410    &0.0387    &0.0375\\
					& CTT(Dec L=1)&0.0346    &0.0355   &0.0363     &0.0381    &0.0385     &0.0393    &0.0396\\
					& CTT(Dec L=2)&0.0348    &0.0359   &0.0365     &0.0375    &0.0391     &0.0412    &0.0406\\
					& CTT(Dec L=3)&0.0348    &0.0361   &0.0365     &0.0383    &0.0391     &0.0400    &0.0432 \\
					& CTT(Dec L=4)&0.0379    &0.0363   &0.0373     &0.0383    &0.0404     &0.0416    &0.0451\\         
					\bottomrule
			\end{tabular}}
		\end{center}
	\end{table*}
	
	\begin{table*}
		\begin{center}
			\caption{RSE, run-time, and communication cost of CTT, with varying $R_1$ and $L$ ($K=4$, 3rd-order synthetic data).}\label{tab:R1_syn}
			{\begin{tabular}{lcccccccc}
					\toprule
					\specialrule{0pt}{2pt}{2pt}
					\multirow{2}{*}{Metric} &\multirow{2}{*}{Algorithms} & \multicolumn{7}{c}{TT rank $R_1$} \\ 
					\cline{3-9} 
					\specialrule{0pt}{2pt}{2pt}
					\multicolumn{2}{c}{}  &5  & 7    &10 &12 & 15 &18 &20  \\ 
					\midrule
					\multirow{5}{*}{RSE}  
					& CTT(M-s) &0.1912  & 0.1867    &0.1832 &0.1820 & 0.1805 &0.1799 &0.1798  \\
					& CTT(Dec L=1) & 0.3390  & 0.3498   & 0.3587  & 0.3620  &0.3631 &0.3644 &0.3649     \\
					& CTT(Dec L=2) & 0.2073  & 0.2048 & 0.2027  & 0.2019   &0.2002 &0.1999 &0.1998    \\
					& CTT(Dec L=3) &0.1929  & 0.1886    &0.1853 &0.1841 & 0.1825 &0.1820 &0.1819  \\
					& CTT(Dec L=4) &0.1914  & 0.1869 & 0.1835  & 0.1822  &0.1807 &0.1801 &0.1800     \\
					\midrule
					\multirow{5}{*}{Communication Cost} 
					& CTT(M-s)& 2.70e+03 & 3.60e+03 & 4.95e+03  & 5.85e+03  &7.68e+03 &9.12e+03 &1.10e+04     \\
					& CTT(Dec L=1) &4.50e+03  & 6.30e+03    &9.00e+03 &1.08e+04 & 1.35e+04 &1.62e+04 &1.80e+04  \\
					& CTT(Dec L=2) & 9.00e+03 &1.26e+04 & 1.80e+04  &2.16e+04   &2.70e+04 &3.24e+04 &3.60e+04    \\
					& CTT(Dec L=3)      &1.35e+04  & 1.89e+04    &2.70e+04 &3.24e+04 & 4.05e+04 &4.86e+04 &5.40e+04  \\
					& CTT(Dec L=4) &1.80e+04       &2.52e+04   &3.60e+04   &4.32e+04 &5.40e+04 &6.48e+04 &7.20e+04 \\
					\midrule
					\multirow{5}{*}{CPU time (seconds)} 
					& CTT(M-s)  &0.0034    &0.0042   &0.0056    &0.0058  &0.0063    &0.0075     &0.0084  \\
					& CTT(Dec L=1) &0.0036    &0.0043   &0.0057    &0.0068   &0.0070   &0.0071     &0.0076 \\ 
					& CTT(Dec L=2) &0.0041    &0.0044   &0.0057    &0.0069   &0.0073   &0.0073     &0.0080 \\
					& CTT(Dec L=3) &0.0045    &0.0054   &0.0062   &0.0070   &0.0073   &0.0078    &0.0081 \\ 
					& CTT(Dec L=4)&0.0040    &0.0062    &0.0067   &0.0071   &0.0077   &0.0079     &0.0080  \\ 
					\bottomrule
			\end{tabular}}
		\end{center}
	\end{table*}
	
	\subsection{Baselines}
	
	We have used the following schemes as comparison baselines.
	\begin{itemize}
		\item D-PSGD~\cite{lian2017can,koloskova2020unified}, a decentralized version that uses SGD to update the factor matrix.
		\item FedGTF-EF~\cite{ma2021communication1}, a master-slave network-based method for communication-effective tensor factorization using fast stochastic gradient compression and error feedback. 
		\item DPFact~\cite{ma2019privacy}, a privacy protection method based on the master-slave network using centralized difference.
	\end{itemize} 
	
	\subsection{Results and Discussion}
	
	\subsubsection{Accuracy}
	
\begin{figure}	\centerline{\includegraphics[width=1.0\columnwidth]{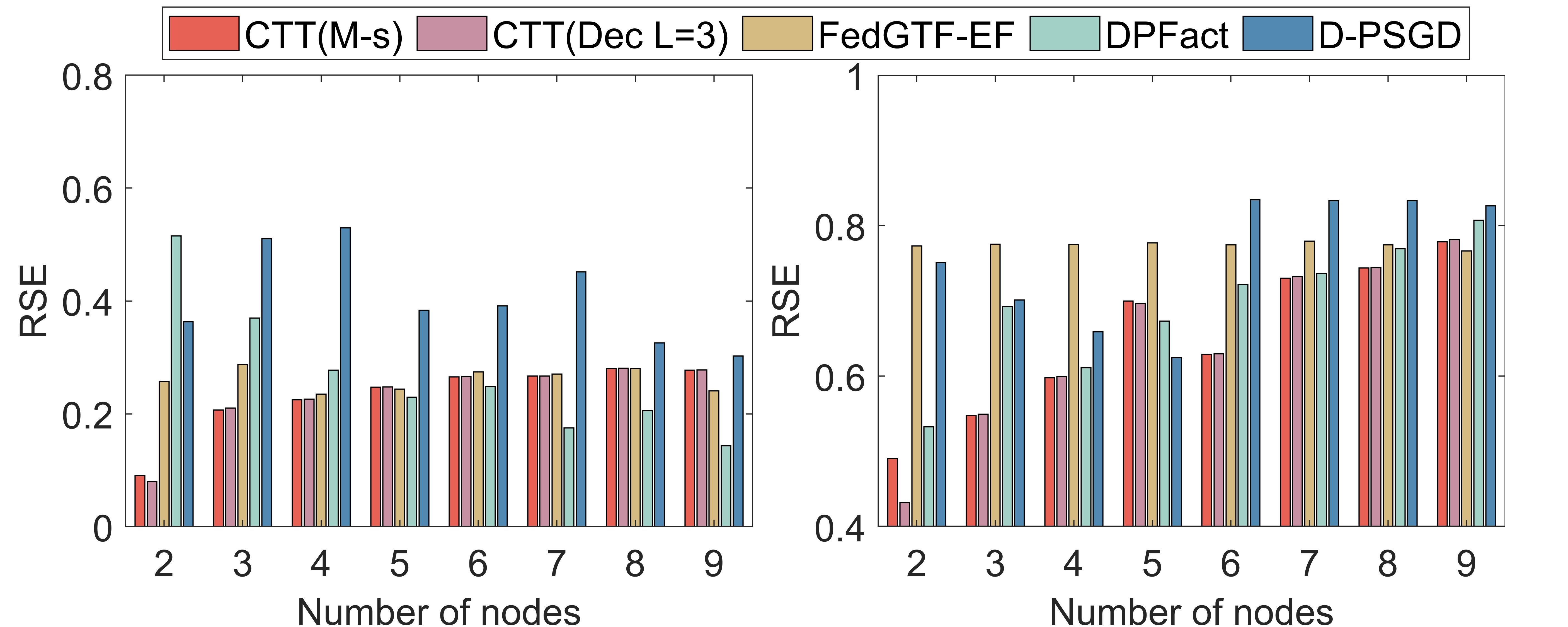}}
		\caption{RSE as a function of the number of nodes, for the Diabetes (left) and ECG (right) datasets.}
		\label{fig:fig_1}
	\end{figure}

	We assess the accuracy of different algorithms by RSE. The obtained results are presented in Fig.~\ref{fig:fig_1} and Table~\ref{tab:compare}.
	\begin{table}
		\begin{center}
			\caption{Comparison of communication cost, run-time, and RSE of different algorithms, on the Diabetes, ECG, and 3rd-order synthetic datasets, with 4 nodes.}
			\label{tab:compare}
			\begin{tabular}{lllll}
				\hline
				{Dataset}	&  {Models} & Rounds & CPU time & RSE  \\ 
				\hline
				\multirow{5}{*}{Diabetes}              
				& FedGTF-EF    & 45    &  0.6250& 0.2347 \\ 
				& D-PSGD    &15    & 0.1641 &0.5294\\ 
				& DPFact  &10     &  0.1480 & 0.2772\\ 
				&CTT(M-s) & \textbf{2} & 0.0410&\textbf{0.2250}\\ 
				&CTT(Dec L=3)  &3  &\textbf{0.0391} & 0.2261 \\ 
				\hline
				\multirow{5}{*}{ECG}                         
				& FedGTF-EF    & 75    &  22.3593& 0.7749 \\
				& D-PSGD    &75    & 22.4023 &0.6589\\
				& DPFact  &10     &  1.3281 & 0.6112\\
				&CTT(M-s) & \textbf{2} &\textbf{0.6117} &\textbf{0.5979}\\
				&CTT(Dec L=3)  &3 &0.6563 & 0.5993 \\
				\hline
				\multirow{5}{*}{Synthetic}                         
				& FedGTF-EF    & 50    & 0.5625& 0.1873 \\
				& D-PSGD    &45    & 0.5469 & \textbf{0.1787}\\
				& DPFact  &6     &  0.4023 & 0.2529\\
				&CTT(M-s) & \textbf{2}   & 0.0084 &0.1798\\
				&CTT(Dec L=3)  &3 &\textbf{0.0081} & 0.1819 \\
				\hline
			\end{tabular}
		\end{center}
	\end{table}CTT achieves higher accuracy in most cases. It is worth noting that when the number of nodes is relatively small, the accuracy advantage of CTT is more obvious. 
	% It is worth noting that it is superior in the master-slave network, which can be attributed to the absence of consensus error in that case. 
	
	\subsubsection{Computational Efficiency}
	
\begin{figure}
		\centerline{\includegraphics[width=1.0\columnwidth]{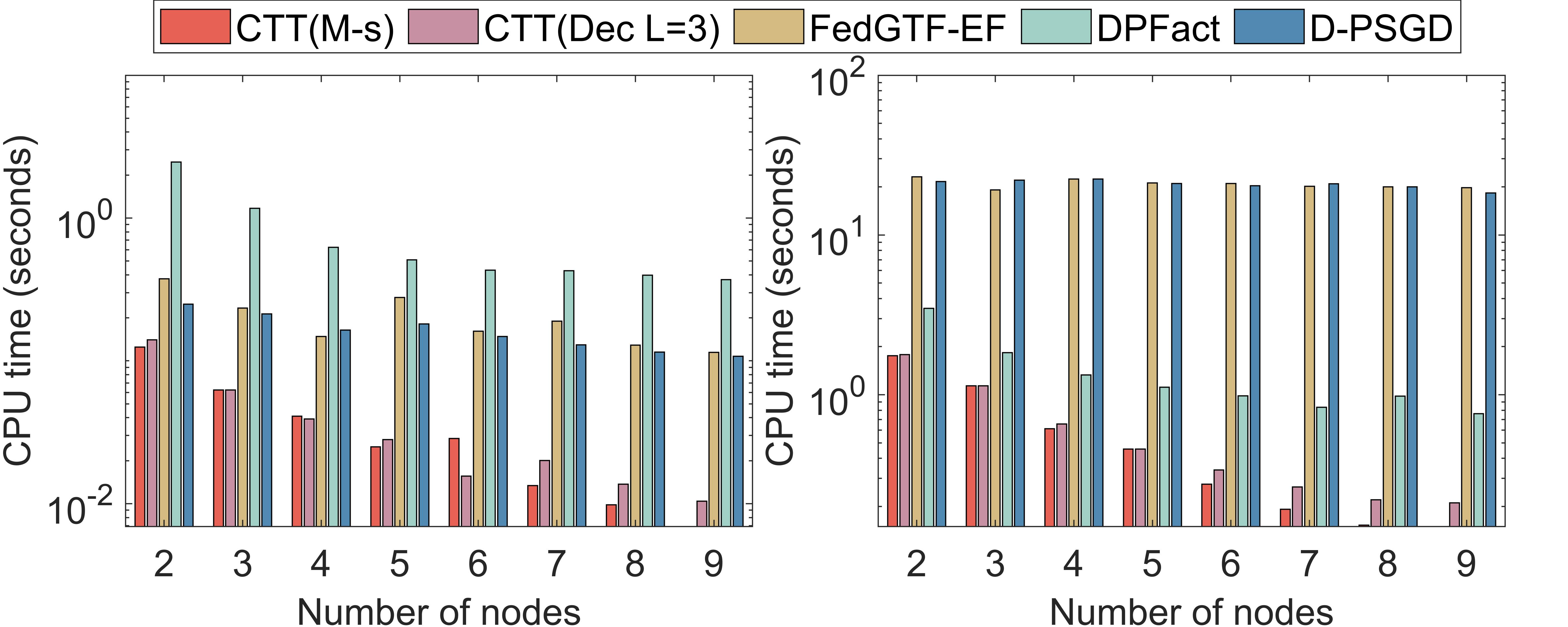}}
		\caption{CPU time as a function of the number of nodes, for the	Diabetes (left) and ECG (right) datasets.}
		\label{fig:com_sumtime}
	\end{figure}

	As shown in Fig.~\ref{fig:com_sumtime}, for the real data examples, CTT outperforms the other algorithms in terms of computational load.  
	The average run-time of 20 repeated experiments is depicted. Note that, for the ECG dataset, the FedGTF-EF and D-PSGD algorithms reached the maximum number of iterations, hence their significantly higher computational cost in that case. Table~\ref{tab:compare} provides additional evidence of the superior computational efficiency of CTT. Notably, this surpasses that of the best baseline method, achieving improvements of 98\%, 54\%, and 74\% with the synthetic, ECG, and Diabetes datasets, respectively.
	
	\subsubsection{Communication Rounds}
	
	The comparison is conducted on the 3rd-order synthetic, ECG, and Diabetes datasets. DPFact, which is only applicable for 3rd-order tensors, is included in the comparison. The results, shown in Table~\ref{tab:compare}, reveal that the master-slave version of our method requires only two communication rounds. This is due to no iterations between the server and the clients are necessary. 
	
	\subsubsection{The Impact of Missing Data}
	
\begin{figure}
		\centerline{\includegraphics[width=1.0\columnwidth]{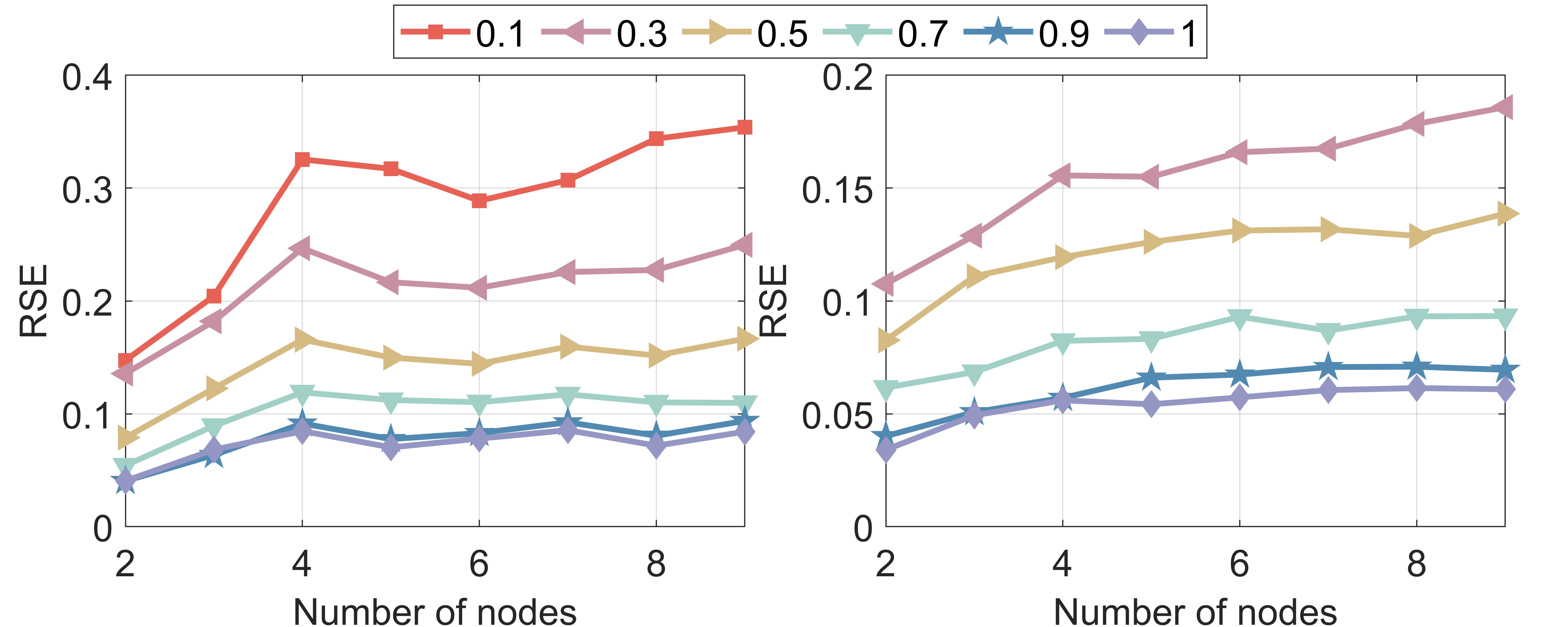}}
		\caption{Effect of the percentage of non-zero entries on the RSE of CTT (M-s), with synthetic data of order~3 (left) and~4 (right). Distinct lines denote varying ratios of non-zero entries.}
		\label{fig:sparsity}
	\end{figure}

	We have also evaluated our method in the presence of missing data entries. We tested this with the synthetic 3rd-order data of dimensions $200\times30\times30$. The percentage of missing data varied from zero to 90\%. The results are shown in Fig.~\ref{fig:sparsity}, for varying numbers of nodes. As expected, CTT exhibits better performance with a diminishing percentage of missing observations. Moreover, the RSE slightly increases with the number of nodes.
	
	\subsubsection{Influence of the Precision Parameter $\epsilon_1$}
	
\begin{figure}
		\centerline{\includegraphics[width=1.0\columnwidth]{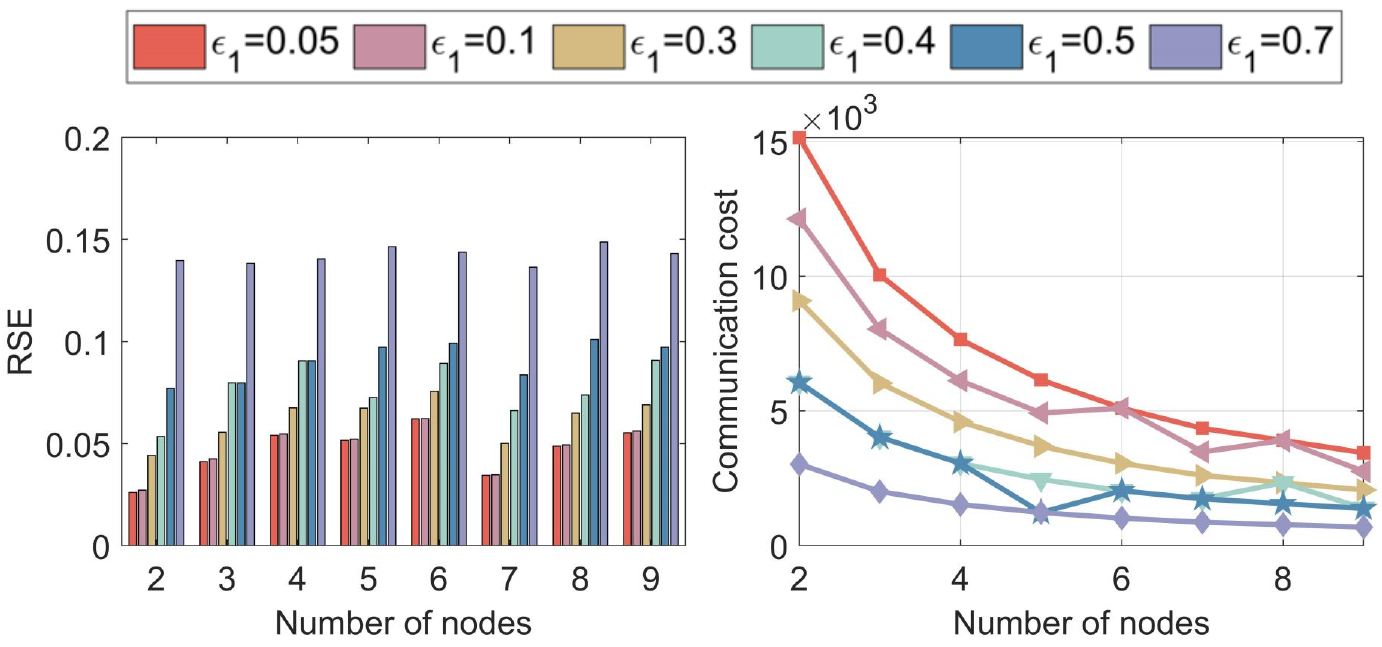}}
		\caption{Effects of different values of $\epsilon_1$ (0.05, 0.1, 0.3, 0.5, 0.7) on the RSE  (left) and the communication cost per link (right) of CTT (M-s), for the synthetic 3rd-order dataset.}
		\label{fig:e1}
	\end{figure}

	To examine the impact of the precision parameter $\epsilon_1$ on the RSE and the communication cost, we utilized $200\times 30\times 30$ synthetic tensors and tested five different values for $\epsilon_1\in\{0.05, 0.1, 0.3, 0.5, 0.7\}$. The results are shown in Fig.~\ref{fig:e1} with a varying number of nodes. 
	As expected, a reduction of $\epsilon_1$ leads to a decrease in the RSE and an increase in the communication cost per link. However, for $\epsilon_1$ less than 0.1, the reduction of RSE is insignificant. Consequently, in other simulation experiments with this dataset, we set $\epsilon_1$ to~0.1.
	
	\subsubsection{Scalability}
	
\begin{figure*}
		\centerline{\includegraphics[width=2.48\columnwidth]{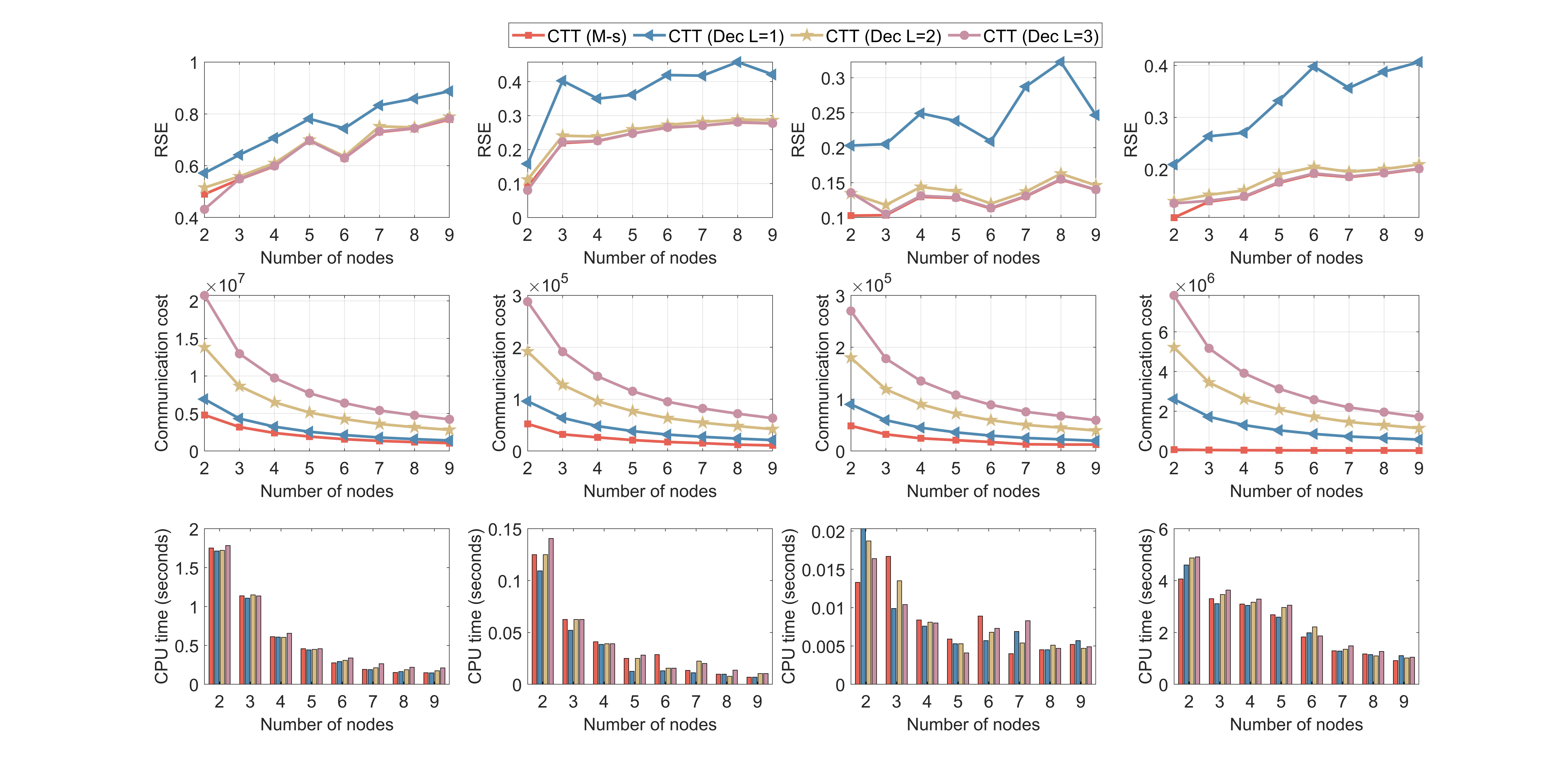}}
		\caption{RSE, communication cost per link, and run-time for CTT versus the number of nodes of different network structures and consensus times when accuracy is optimal.
			The results correspond to the ECG, Diabetes, 3rd-order, and 4th-order synthetic data, respectively, from left to right.}
		\label{fig:scalability}
	\end{figure*}

	We assess the horizontal scalability of the proposed method by systematically increasing the number of nodes, where the number of nodes represents the network size, and examining the corresponding changes in RSE, communication cost, and computational efficiency. Analyzing the results depicted in Fig.~\ref{fig:scalability}, we observe that as the number of nodes increases, there is a slight increase in the RSE of CTT, indicating a slight degradation in accuracy. 
	However, the total run-time exhibits a significant decrease, demonstrating the improved computational efficiency achieved in a distributed setting. Furthermore, the second row of Fig.~\ref{fig:scalability} illustrates that, as the number of nodes increases, the communication cost per link decreases, resulting in improved communication efficiency. Of course, achieving optimal performance in practice requires careful optimization of the fundamental trade-off between computational and communication efficiency.
	
	\subsubsection{The Impact of the Topology}
	
	We have also studied the effect of the network topology on the RSE and the total communication cost of all links with CTT. Specifically, we considered the master-slave network as well as decentralized networks with varying densities. The latter is quantified by the ratio of the actual number of links over the number of links in the corresponding fully connected graph, denoted by $S$. 
	
\begin{figure}
		\centerline{\includegraphics[width=1.0\columnwidth]{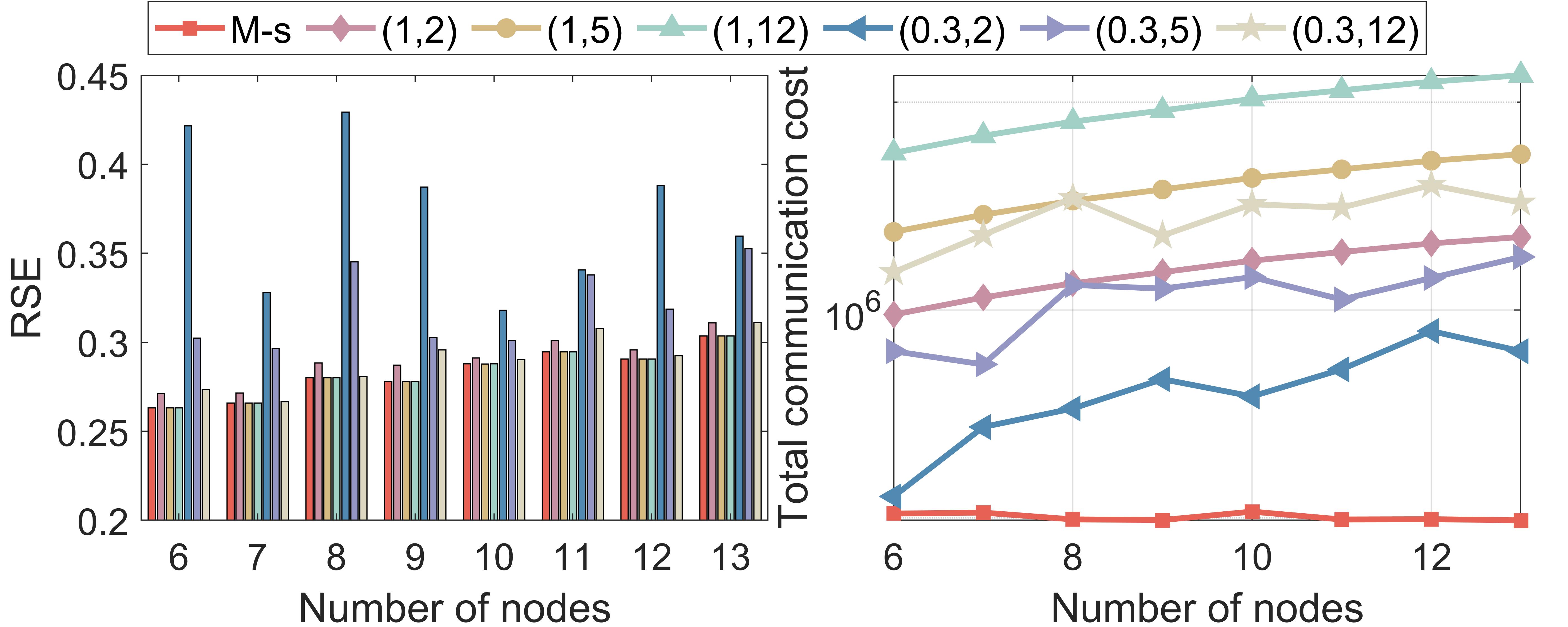}}
		\caption{The RSE and total communication cost of CTT in different network structures (master-slave (M-s), decentralized ($S,L$) with connectivity $S$ and consensus times $L$), for the Diabetes dataset.}
		\label{fig:graph}
	\end{figure}

	The first row of Fig.~\ref{fig:scalability} shows that a fully connected decentralized network only requires $L=3$ consensus iterations to achieve the minimum error. For networks of lower connectivity, higher $L$ values are needed, as exemplified in Fig.~\ref{fig:graph} (left). 
	Furthermore, from Fig.~\ref{fig:graph} (right), we conclude that the total communication cost of a decentralized network with low connectivity is lower. Consequently, in practical scenarios, selecting an appropriate connectivity level for a decentralized network becomes crucial, particularly when the number of nodes is pre-determined.
	
	\subsubsection{Classification Accuracy}
	
\begin{figure}
		\centerline{\includegraphics[width=1.0\columnwidth]{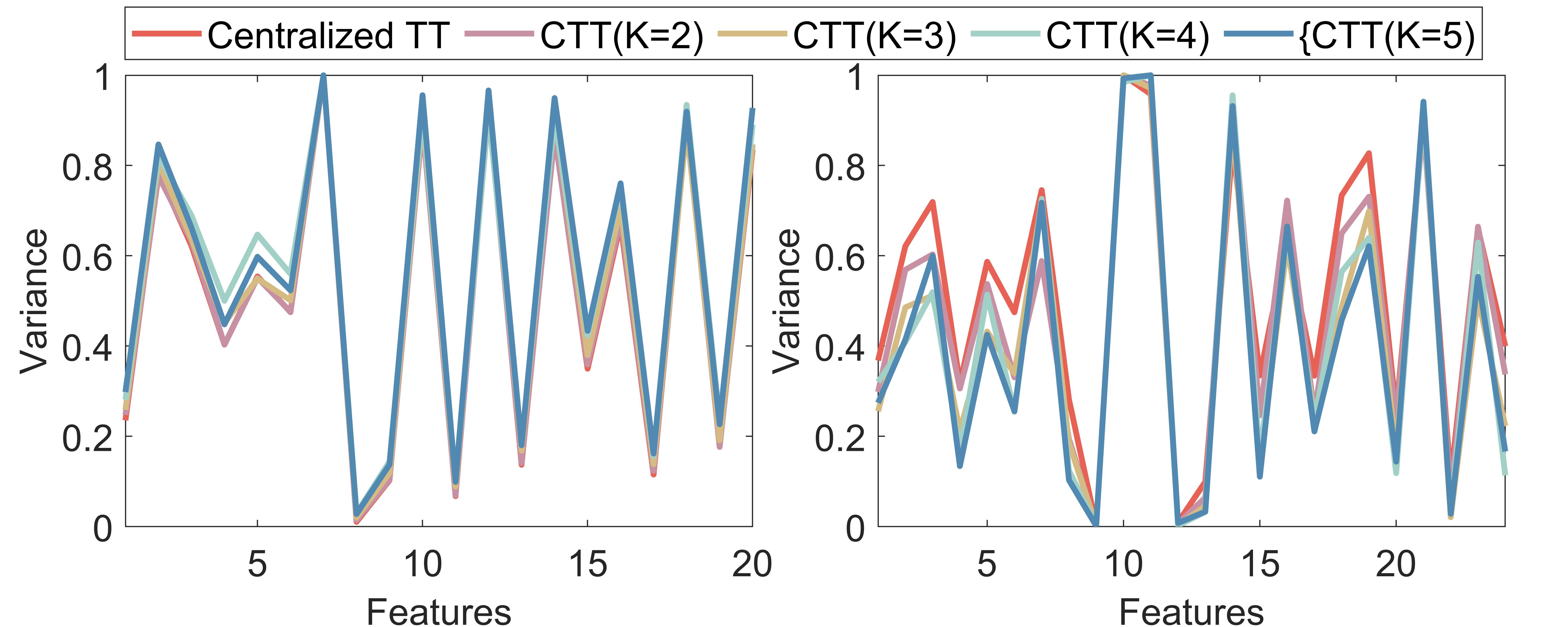}}
		\caption{The variance of the features of the second mode (left) and the third mode (right) in the Diabetes dataset ($K$ stands for the number of nodes in the CTT (M-s) configuration).}
		\label{fig:feature_var}
	\end{figure}

\begin{figure}
		\centerline{\includegraphics[width=1.0\columnwidth]{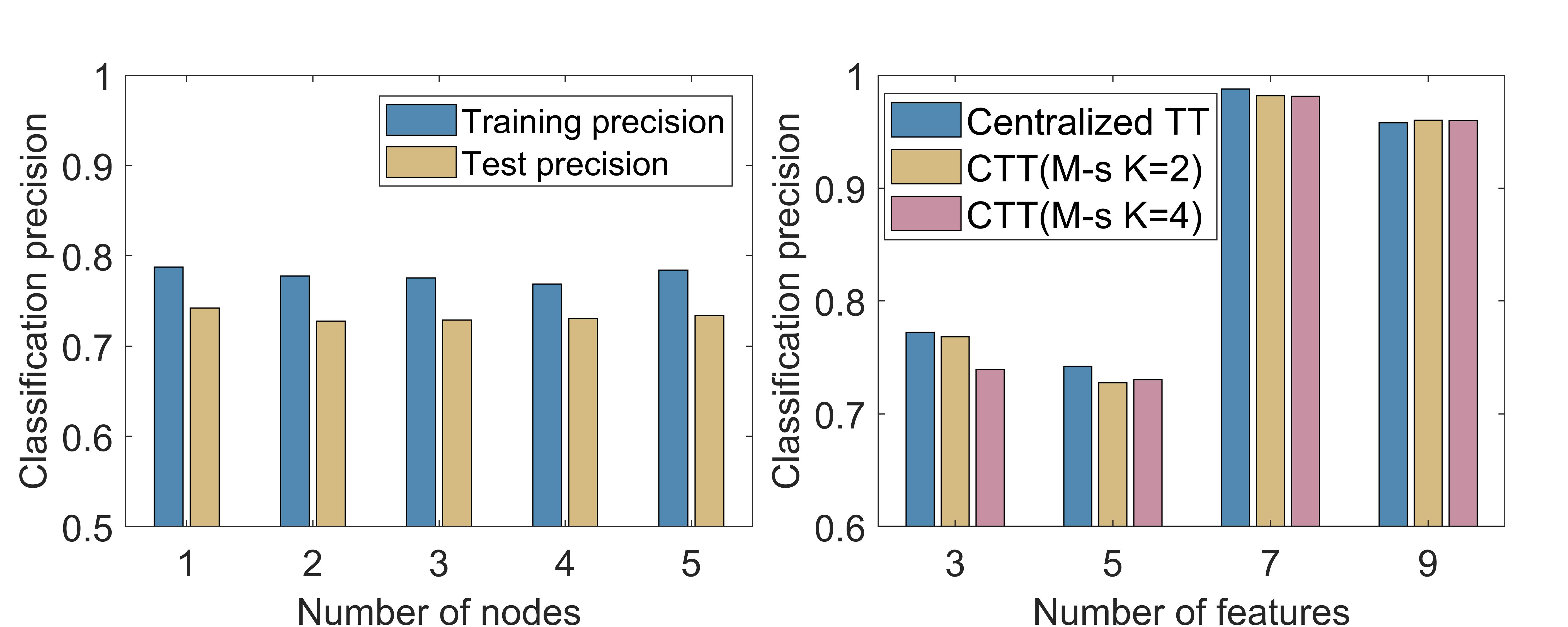}}
		\caption{Left: Training accuracy and test accuracy as functions of the number of nodes, with $m=5$ selected features. Right: Test accuracy as the number of selected features, $m$, varies.}
		\label{fig:feature_classify}
	\end{figure}

	The Diabetes data are categorized as follows: 0 designates cases with no diabetes or only during pregnancy, 1 corresponds to pre-diabetic condition, and 2 pertains to diabetes. We built a classifier for this dataset using the feature model TT cores as features, as follows. For the $n$th feature mode, there are $I_n$ features of dimension $R_{n-1}R_{n}$, $n=2,\ldots,N$. Their variances are computed and we select the $m$ features with the highest variance, with $m$ being user-chosen. We have adopted the k-nearest neighbor (kNN) classification model to perform category prediction using the selected features. Training and test datasets are chosen to maintain a ratio of 7:3 in their sizes. The classification accuracy is computed by averaging the outcomes of ten cross-validation experiments. We compared master-slave CTT with the Centralized TT approach, wherein all data are available at the server before TTD. Fig.~\ref{fig:feature_var} shows that the extracted representative global features from multiple nodes closely align with the representative features extracted from the centrally decomposed data. 
	As the right-hand side of Fig.~\ref{fig:feature_classify} demonstrates, the CTT approach attains a classification accuracy comparable with its centralized counterpart, for several values of $m$.
	Moreover, as one can see on the left-hand side of Fig.~\ref{fig:feature_classify}, the CTT performance is effectively independent of the network size with at least $m=5$ features.
	
	\section{Conclusion}
	\label{sec:conclusion}
	
	To the best of our knowledge, this is the first time that a federated tensor decomposition approach has been developed based on the TT model and its coupled version. In the proposed so-called CTT framework, we presented two such schemes, applicable in master-slave and decentralized network structures, and analyzed their computation and communication requirements and their abilities for privacy preservation. Simulation experiments, with both synthetic and real data, have been reported that demonstrate the superiority of the proposed method over existing related ones relying on CPD. Figures of merit employed in the comparative study include accuracy of decomposition and computational and communication efficiency. Furthermore, CTT performance was evaluated for varying network topologies, sizes, and densities, as well as percentages of missing data. A result worth strongly emphasizing is that the loss in feature extraction performance over a centralized and non-FL TT method is negligible, as demonstrated with the aid of a classification experiment with real medical data. This implies that the proposed method achieves all the objectives of an FL environment, while outperforming existing alternatives and with practically no loss in learning performance incurred from its distributed character. 
	
	In addition to further experimentation, future work should include investigating ways of overcoming the requirement of all $R_1^k$ being equal. 
	
	\bibliographystyle{IEEEtran}
	\bibliography{IEEEabrv,refs}

\end{document}